\documentclass[10pt, conference, compsocconf]{IEEEtran}

\usepackage{tikz}
\usepackage{fancyhdr}
\usepackage{amsfonts, amsmath, amssymb}
\usepackage{booktabs}
\usepackage{multirow}
\usepackage{graphicx, cite, xcolor, setspace, url, verbatim, afterpage}
\usepackage{stfloats}
\usepackage{xspace}
\usepackage{paralist}
\usepackage{textcomp}
\usepackage{listings}
\usepackage{algorithm}
\usepackage[noend]{algpseudocode}

\usepackage{soul,color}

\newcommand{\secref}[1]{Section~\ref{#1}}

\newcommand{\figref}[1]{Figure~\ref{#1}}

\newcommand{\tabref}[1]{Table~\ref{#1}}
\newcommand{\ignore}[1]{}

\def\naive{na\"\i ve}

\def\doorbell{\emph{DoorBell}}
\def\doorbells{\emph{DoorBells}}
\def\postlist{\emph{Postlist}}
\def\inlining{\emph{Inlining}}
\def\unsig{\emph{Unsignaled Completion}}
\def\unsigs{\emph{Unsignaled Completions}}
\def\blueflame{\emph{BlueFlame}}

\lstdefinestyle{customc}{
	belowcaptionskip=1\baselineskip,
	breaklines=true,
	frame=L,
	xleftmargin=\parindent,
	language=C,
	tabsize=1,
	numbers=left,
	showstringspaces=false,
	basicstyle=\footnotesize\ttfamily,
	keywordstyle=\bfseries\color{green!40!black},
	commentstyle=\itshape\color{purple!40!black},
	identifierstyle=\color{blue},
	stringstyle=\color{orange},
}

\begin{document}

\title{Scalable Communication Endpoints for MPI+Threads Applications}

\author{Rohit Zambre,$^{\star}$ Aparna Chandramowlishwaran,$^{\star}$ Pavan Balaji$^{\dagger}$ 
	\vspace{0.1in}\\
	{\em  $^{\star}$EECS, University of California, Irvine, CA}\\
	{\em  $^{\dagger}$MCS, Argonne National Laboratory, Lemont, IL} \\
}
\IEEEoverridecommandlockouts
\maketitle
\begin{abstract}

  Hybrid MPI+threads programming is gaining prominence as an
  alternative to the traditional ``MPI everywhere'' model to better
  handle the disproportionate increase in the number of cores compared
  with other on-node resources. Current implementations of these two
  models represent the two extreme
  cases of communication resource sharing in modern MPI
  implementations. In the MPI-everywhere model, each MPI process has a
  dedicated set of communication resources (also known as
  endpoints), which is ideal for performance but is
  resource wasteful. With MPI+threads, current MPI implementations
  share a single communication endpoint for all threads, which is
  ideal for resource usage but is hurtful for performance.

  In this paper, we explore the tradeoff space between performance and
  communication resource usage in MPI+threads environments. We first
  demonstrate the two extreme cases---one where all threads share a
  single communication endpoint and another where each thread gets its
  own dedicated communication endpoint (similar to the MPI-everywhere
  model) and showcase the inefficiencies in both these cases. Next,
  we perform a thorough analysis of the different levels of resource
  sharing in the context of Mellanox InfiniBand. Using the
  lessons learned from this analysis, we design an improved 
  resource-sharing model to produce \emph{scalable communication endpoints} that can
  achieve the same performance as with dedicated
  communication resources per thread but using just a third of the
  resources.

\end{abstract}

\vspace{-1em}
\section{Introduction}
\label{sec:introduction}

The Message-Passing Interface (MPI) is the most commonly used model
for programming large-scale parallel systems today.  The traditional
model for using MPI hitherto has been the ``MPI everywhere"
model in which the application launches an MPI process on each core
of the supercomputer and executes by ignoring the fact that some of
the MPI processes reside on different cores of the same node while
some execute on different nodes. The MPI implementation then
internally optimizes communication within the node by using shared
memory or other techniques.

\begin{figure}[htbp]
	\vspace{-1em}
	\begin{center}
		\includegraphics[width=0.49\textwidth]{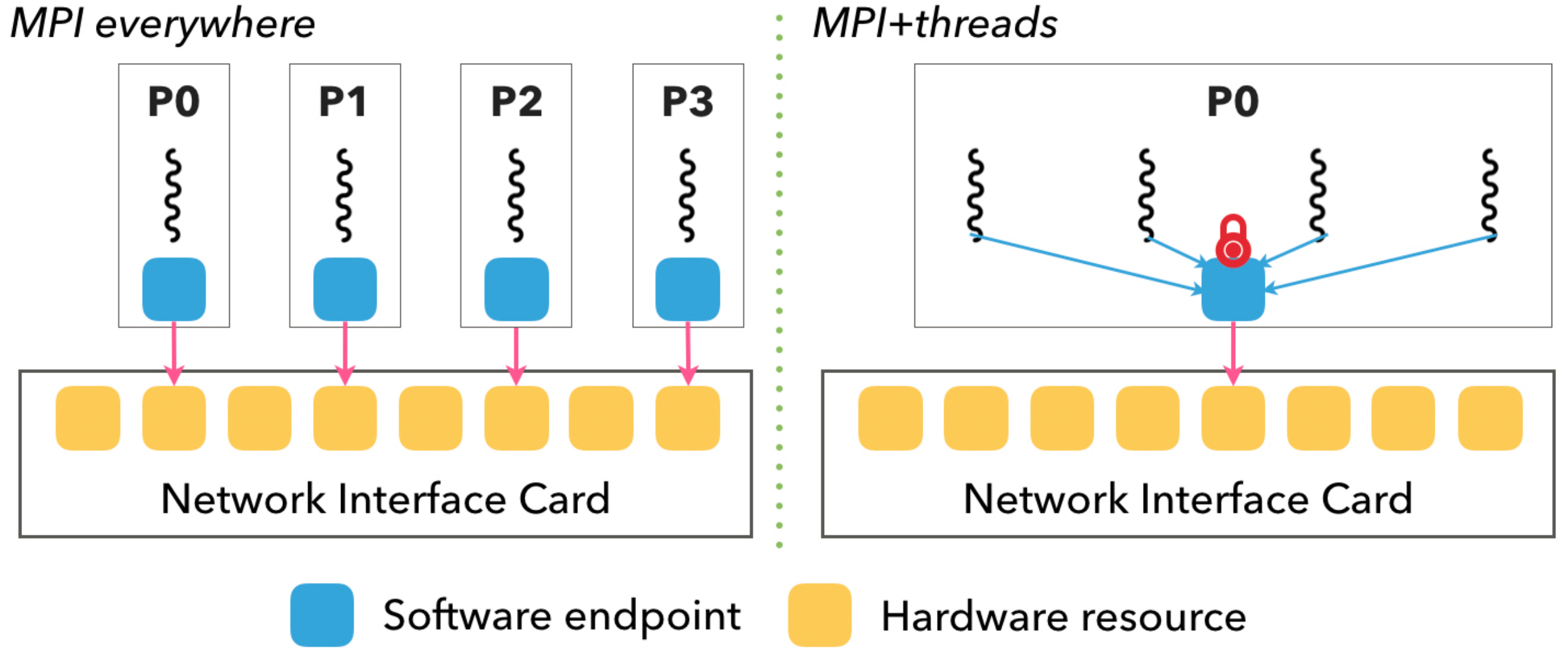}
	\end{center}
	\vspace{-1em}
	\caption{Endpoint configuration of MPI-everywhere and MPI+threads
		models.}
	\label{fig:sotaeps}
	\vspace{-1em}
\end{figure}

While the MPI-everywhere model of parallelism has served applications
well for several decades, 
scaling applications in this model is becoming increasingly difficult. The biggest reason for this
difficulty in scaling is that not all on-node resources scale
at the same rate.  Specifically, the number of cores available on a
node is increasing rapidly.  Other on-node resources such as memory,
cache, TLB space, and network resources, however, scale much more
slowly.  Since the MPI-everywhere model uses a separate MPI process
for each core, it inadvertently leads to a static split of all on-node
resources, resulting in underutilization and wastage of resources.  While
optimizations such as \emph{MPI shared memory}~\cite{hoefler2013mpi+}
address sharing a subset of resources (in particular, memory), these
optimizations are not a generic solution for all on-node resources.
Consequently, researchers have been increasingly looking at hybrid
MPI+threads programming (e.g., MPI+OpenMP) as an alternative to the
traditional MPI-everywhere model~\cite{thakur2010mpi}.

Current implementations of these two models---MPI everywhere and
MPI+threads---represent the two extreme cases of communication
resource sharing in modern MPI implementations.  \figref{fig:sotaeps}
contrasts these two models in state-of-the-art MPI implementations,
such as MPICH~\cite{mpich}, that use one communication endpoint per
MPI process~\cite{thakur2010mpi}.  A communication endpoint is a set
of communication resources that allows the software to interface with
the network hardware to send messages over the network.

In the MPI-everywhere model, multiple communication endpoints exist
per node where each MPI process communicates using its own endpoint.
This allows each MPI process to communicate completely independently
of other processes, thus providing a direct and contention-free path
to the network hardware and leading to the best-achievable
communication performance (assuming that the MPI implementation is
sufficiently optimized).  In the MPI+threads model, on the other hand,
all threads within an MPI process communicate using a single endpoint, which 
causes the MPI implementation to use locks on the endpoint for serialization.  
This model hurts communication throughput; more important, the available
network-level parallelism remains underutilized.  But, this model 
uses the least possible amount of communication resources.

\begin{figure*}[htbp]
    \centering
    \begin{minipage}{0.49\textwidth}
      \centering
      \includegraphics[width=\textwidth]{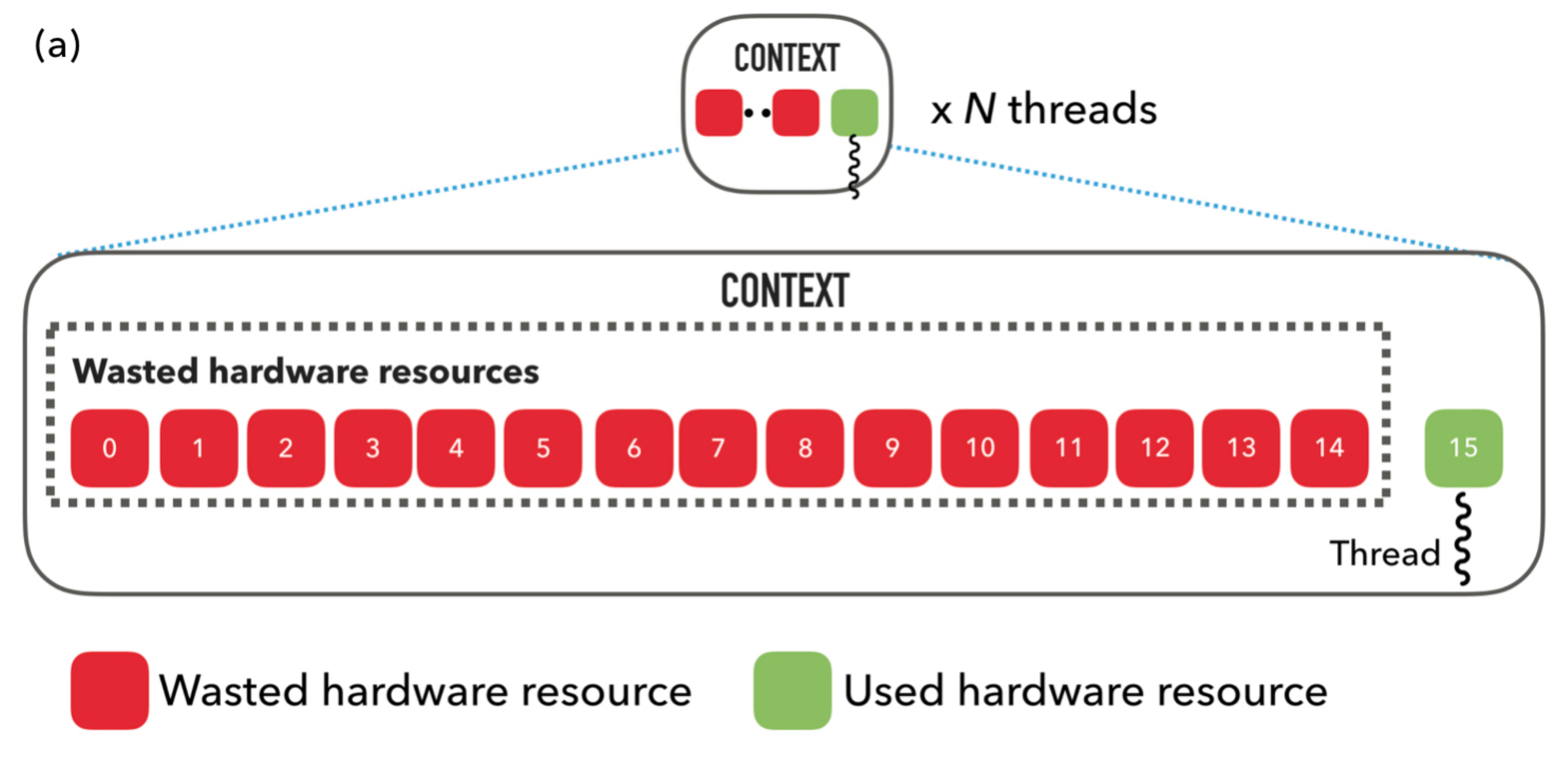}
      \vspace{-1.5em}
    \end{minipage}
    \begin{minipage}{0.49\textwidth}
      \centering
      \includegraphics[width=\textwidth]{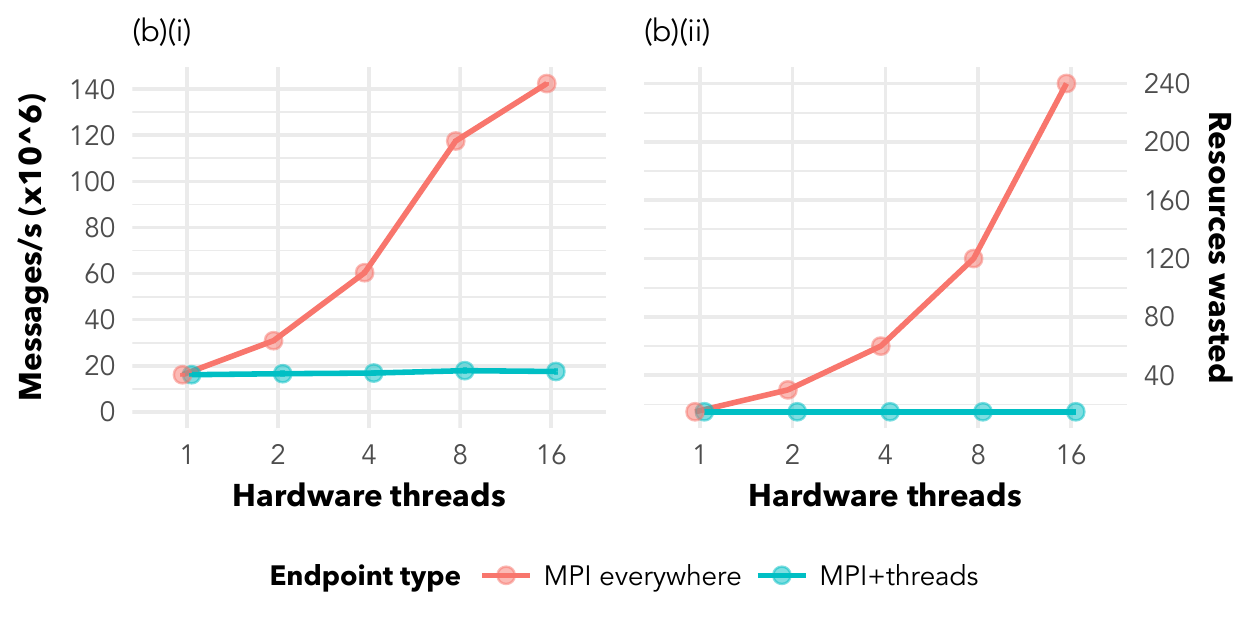}
      \vspace{-1.5em}
    \end{minipage}    
    \caption{(a) Demonstration of 93.75\% hardware resource wastage
      per context in the \naive\ approach. (b) (i) Throughput (higher
      is better) and (ii) number of wasted hardware resources (lower
      is better) with state-of-the-art endpoints on Mellanox's
      ConnectX-4 adapter.}
    \label{fig:tputvswastage}
    \vspace{-1em}
\end{figure*}

A straightforward way to achieve maximum communication path
independence between threads in the MPI+threads model is to dedicate a
separate context each containing an endpoint with its own set of
resources to each thread.  This emulates the endpoint configuration
in the MPI-everywhere model where each MPI process has its own
context.  Although such a \naive\ approach can achieve the maximum
throughput for a given number of threads, it wastes the hardware's
limited resources.  \figref{fig:tputvswastage}(a) shows how this
\naive\ approach translates to 93.75\% hardware resource wastage on a
modern Mellanox mlx5 InfiniBand device.  In order to achieve maximum
resource efficiency, multiple threads can share just one endpoint,
which is the case for the MPI+threads model in state-of-the-art MPI
implementations.  Doing so, however, drastically impacts communication
throughput.  \figref{fig:tputvswastage}(b) shows the tradeoff between
throughput and hardware resource wastage in a multithreaded
environment that emulates state-of-the-art endpoints in the
MPI-everywhere and MPI+threads models.

Note that the MPI+threads model itself does not force the extreme of 
using a single endpoint for communication by all threads.  That is simply 
how the state-of-the-art MPI libraries implement it.  Unlike 
MPI everywhere, the MPI+threads environment allows for any arbitrary level
of sharing of communication resources between the different threads.
The question that we really need to answer is, \textit{what level of
  resource sharing is ideal?}  As is the case with any computer
science question that is worth its salt, the answer is, \textit{it depends.}  
If one is looking for the least amount of resources to
use without losing any performance compared with the MPI-everywhere
model, a certain set of resources can be shared while others cannot.  
If a small percentage of performance loss is acceptable, a
different division of shared vs. dedicated resources would be ideal.
If resource efficiency is the most important criterion and additional
performance loss is acceptable, yet another division of shared
vs. dedicated resources would be ideal.  Understanding this tradeoff
space between performance and resource usage is the primary goal of
this paper.

To that end, this paper makes the following contributions.
\begin{enumerate}
\item We demonstrate the two extreme cases---one where all
  threads share a single communication endpoint and another
  where each thread gets its own dedicated endpoint. We showcase 
  the inefficiencies in both these cases.
\item We explore the tradeoff space between performance
  (communication throughput) and communication resource
  usage in a multithreaded environment.
  In \secref{sec:resources},
  we first discuss
  the communication resources of an endpoint. 
In \secref{sec:analysis}, we
  thoroughly analyze the different levels of resource sharing
  in MPI+threads environments in the context of Mellanox
  InfiniBand, the most popular high-speed interconnect on
  the TOP500 and also the preferred interconnect for both artificial intelligence and high-performance computing (HPC)~\cite{ibtop500}.
\item Using the lessons learned from our analysis, we design
  efficient resource-sharing models in \secref{sec:scalableepdesign}
  to provide \emph{scalable communication endpoints}. Scalable
  endpoints provide a wide range of resource-sharing models,
  ranging from fully independent to fully shared  communication paths.
  Our evaluation on scalable
  endpoints in \secref{sec:scalableepeval} shows that
  fully independent communication paths can achieve performance
  as high as MPI-everywhere endpoints by using 3.2x fewer resources.
\end{enumerate}

\section{Background}
\label{sec:background}

InfiniBand (IB) is the popular choice among high-speed interconnects.
Mellanox
Technologies is the most renowned IB vendor, powering 216 systems
(both IB and Ethernet) on the TOP500~\cite{ibtop500}. Hence, we study
the mlx5 provider of Verbs, the IB software stack. Mellanox's
Connect-IB adapter and its ConnectX series,
starting from ConnectX-4, are mlx5 devices.

\subsection{InfiniBand Resources}
\label{sec:ibresources}

The software bidirectional communication portal in IB is the queue pair (QP):
a pair of send and receive FIFO queues, to which work queue entries (WQEs), IB's
message descriptors, are posted. Each QP is associated with a completion queue
(CQ) that contains completion queue entries (CQEs) corresponding to the completion of signaled WQEs.
To create a QP, we need at least one memory buffer (BUF),
device context (CTX), protection domain (PD), and CQ. A memory region (MR)
is required if the NIC needs direct access to memory. Chapter 10 of
the IB specification details the IB resources~\cite{ibspec}. Additionally,
we can assign QPs to thread domains (TDs) to provide single-threaded access hints
to the QPs in a TD.

The CTX is the container of all IB resources and is also a slice of the network hardware,
containing a subset of the NIC's hardware resources. In mlx5 devices, the hardware
resources are part of the user access region (UAR) of the NIC's address space. Each
UAR page consists of two micro UARs (uUARs). By default, a CTX contains eight UARs (UAR pages) and,
hence, 16 uUARs. The user's QPs are mapped to one of the \emph{statically allocated}
uUARs unless a QP is part of a TD in which case the QP is mapped to a uUAR in a UAR
that was \emph{dynamically allocated} during TD creation. ~\cite{scalable}
details these resources and describe mlx5's uUAR-to-QP assignment policy.

\subsection{InfiniBand Operational Features}
\label{sec:ibfeatures}

To send a message on InfiniBand, the application calls \texttt{ibv\_post\_send}.
What follows is a series of coordinated operations between the CPU and the
NIC to fetch the WQE (DMA read), read its payload (DMA read), and signal its completion (DMA write).
~\cite{scalable} portrays the operations involved.

The NIC is typically a PCIe device and hence, the overhead of the
operations is multiple PCIe round-trip latencies. Naturally, reducing
the number of round-trip latencies for small messages impacts
throughput significantly. \inlining, \postlist, \unsigs, and \blueflame\
are IB's operational features that help reduce this overhead. We describe them
below considering the depth of the QP to be $n$.

\noindent \emph{\textbf{Postlist.}} Instead of posting only one WQE per
\texttt{ibv\_post\_send}, IB allows the application to post a
linked list of WQEs with just one call to \texttt{ibv\_post\_send}. It
can reduce the number of \doorbell\ rings from $n$ to 1.

\noindent \emph{\textbf{Inlining.}} Here, the CPU copies the data
into the WQE. Hence, with its first DMA read for the WQE, the NIC gets
the payload as well, eliminating the second DMA read for the payload.
 
\noindent \emph{\textbf{Unsignaled Completions.}} Instead of signaling
a completion for each WQE, IB allows the application to turn off
completions for WQEs provided that at least one out of every \emph{n}
WQEs is signaled. Turning off completions reduces the DMA writes of CQEs
by the NIC. Additionally, the application polls fewer CQEs, reducing the
overhead of making progress.

\noindent \emph{\textbf{BlueFlame.}} \blueflame\
is Mellanox's terminology for programmed
I/O---it writes the WQE along with the \doorbell, cutting off the first DMA read.
With \blueflame, the UAR pages are mapped as 
write-combining (WC) memory. Hence, the WQEs sent using \blueflame\ are buffered
through the CPU's WC buffers. Note that \blueflame\ is not used with
\postlist; the NIC will DMA-read the WQEs in
the linked list.

Using both \inlining\ and \blueflame\ for small messages eliminates two
PCIe round-trip latencies. While the use of \inlining\ and \blueflame\ is
dependent on message size, the use of \postlist\ and \unsigs\ is
reliant primarily on the user's design choices and application semantics.

\section{Communication Resources}
\label{sec:resources}

To send messages across the network, the software (CPU)
coordinates with the hardware (NIC) to \emph{initiate} a transfer
and confirm its \emph{completion}. This coordination occurs
through three communication resources: a software transmit
queue, a software completion structure, and a NIC's hardware
resource. The three interact using the mechnasims described in
~\cite{scalable} and features described in \secref{sec:ibfeatures}.
In IB, the transmit queue is the QP,
the completion structure is the CQ, and the hardware resource
is the uUAR contained within a UAR page. The QP, UAR, and
uUAR make up the \emph{initiation} interface; the CQ is
the \emph{completion} interface.

The threads of a MPI+threads application eventually map to QPs,
and the QPs eventually map to a uUAR on a UAR of the NIC.
As seen in \secref{sec:ibresources}, the interconnect's
driver dictates the mapping between the transmit queues and the
hardware resources while the user decides the
mapping between the transmit queues and completion structures.
Multiple QPs could share the same CQ, or each could have its own.

The QP and CQ are associated with circular buffers that contain
their WQEs and CQEs, respectively. The CPU writes to the QP's buffer,
and the NIC DMA-reads it when \inlining\ is not used. The NIC
DMA-writes the CQ's buffer and the CPU reads it when polling for progress.
Both buffers are pinned by the operating system during resource creation.

\begin{table}
  \centering
  \caption{Bytes used by mlx5 Verbs resources}
  \label{tab:resmem}
  \begin{tabular}{  c | c | c | c | c | c }
    \hline
    \textbf{CTXs} & \textbf{PDs} & \textbf{MRs} & \textbf{QPs} & \textbf{CQs} 	& \textbf{Total} \\ \hline
    256K & 144 & 144 & 80K & 9K & 345K \\
    \hline
  \end{tabular}
	\vspace{-1.25em}

\end{table} 

The QP and CQ occupy memory with their circular buffers. So,
every time we create a QP or a CQ, we impact memory
consumption. \tabref{tab:resmem} shows the memory used by
each type of a Verbs resource (for mlx5) that is required to
open a QP. Creating one endpoint requires at least 354 KB
of memory, with the CTX occupying 74.2\% of it.

However, the memory usage of the QP and the CQ is on the
order of kilobytes, whereas the memory on the nodes of
clusters and supercomputers is typically on the order of hundreds of
gigabytes. Hence, we will notice a formidable impact on memory consumption
only when the number of the Verbs resources is on the order of thousands.
The impact of creating a QP or a CQ on memory is not of immediate
concern.

On the other hand, the limit on the hardware resource is much smaller:
8K UAR pages on the ConnectX-4 NIC with only two uUARs
per UAR. The situation is similar for other interconnects such as
Intel Omni-Path, where the maximum number of hardware contexts
on its NIC is 160~\cite{hfi_guide}. The 8K UARs on
ConnectX-4 translates to a maximum of 907 CTXs, considering that the
user creates a TD-assigned QP contained within its own CTX for each
thread. Each CTX contains a total of 18 uUARs---the 16 static ones plus
the two from the TD's dynamically allocated UAR (see \secref{sec:ibresources}).
The resource wastage of this approach is a
staggering 94\% since it uses only one uUAR out of 18. Arguably, we will not run out of
hardware resources even if we create one endpoint per core on
existing processors with this
approach, but eliminating this huge
wastage would enable vendors to significantly reduce the power and cost
of their NICs. Such high wastage translates to requiring a
second NIC on the node after only marginally utilizing the resources on the
first.

\section{Evaluation Setup}
\label{sec:setup}

To evaluate the impact of resource sharing
on performance, we write a
multithreaded ``sender-receiver," RDMA-write message rate
benchmark. We choose RDMA writes to
eliminate any receiver-side processing on the
critical path.

We conduct our study on the Joint Laboratory for
System Evaluation's Gomez cluster
(each node has quad-socket Intel Haswell processors with 16 cores/socket and one hardware thread/core) using a patched
rdma-core~\cite{rdmacore} library that contains the infrastructure
to allow for maximally independent paths and disabled
mlx5 locks as described in
\secref{sec:ctxsharing}. The two nodes are
connected via a switch, and each
node hosts a single-port Mellanox ConnectX-4 NIC.
We ensure that each
thread is bound to its own core. For repeatable and
reliable measurements, we disable the processor's
turbo boost and set the CPU frequency to 2.5 GHz.

The design of our message-rate benchmark is
adopted from \texttt{perftest}~\cite{perftest}. 
The loop of a thread iterates until all its messages are
completed. In each iteration, the thread posts WQEs on a QP of depth,
$d$ in multiples of
\postlist, $p$ requesting for one signaled completion
every $q$ WQEs, where $q$ is the value of \unsigs. In
each poll on the CQ, the thread requests for $c = d/ q$
completions, namely, all possible completions in an
iteration. The depth of the
CQ is $c$.

\postlist\ and \unsigs\ control the rate and amount of
interaction between the CPU and NIC.
Empirically, we find that setting $p=32$ and
$q=64$ achieves the maximum throughput for 16 threads; hence, we
use them as our default values. Note that we define the
values of \postlist\ and \unsigs\ with respect to the
threads, not to their associated QPs.

To study the effect of an IB feature, we
remove that feature while using others,
referring to this case as ``All w/o $f$," where $f$ is the
feature of interest. To disable \blueflame, we set the MLX5\_SHUT\_UP\_BF
environment variable. To enable \inlining, we set
the IBV\_SEND\_INLINE flag on the send-WQE. We use ``w/o Postlist" to mean
$p=1$, and similarly ``w/o Unsignaled" to mean $q=1$.

\begin{figure}[htbp]
	\centering
	\begin{minipage}[t]{0.24\textwidth}
		\centering
		\includegraphics[width=\textwidth]{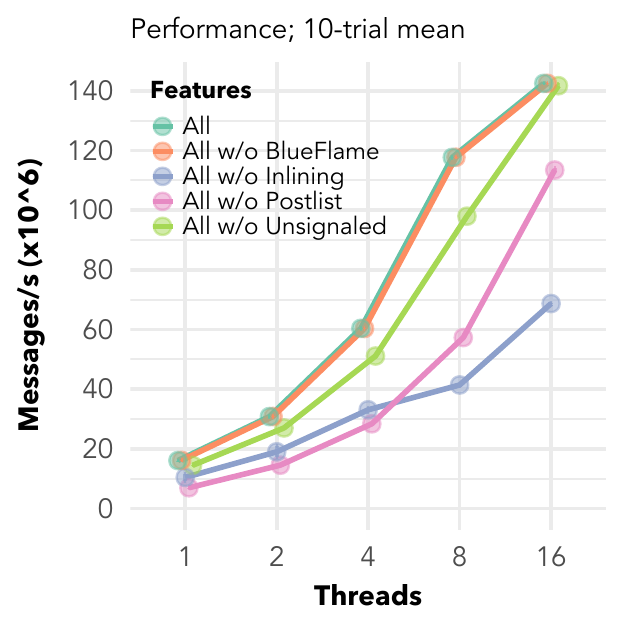}
	\end{minipage}
	\begin{minipage}[t]{0.24\textwidth}
		\centering
		\includegraphics[width=\textwidth]{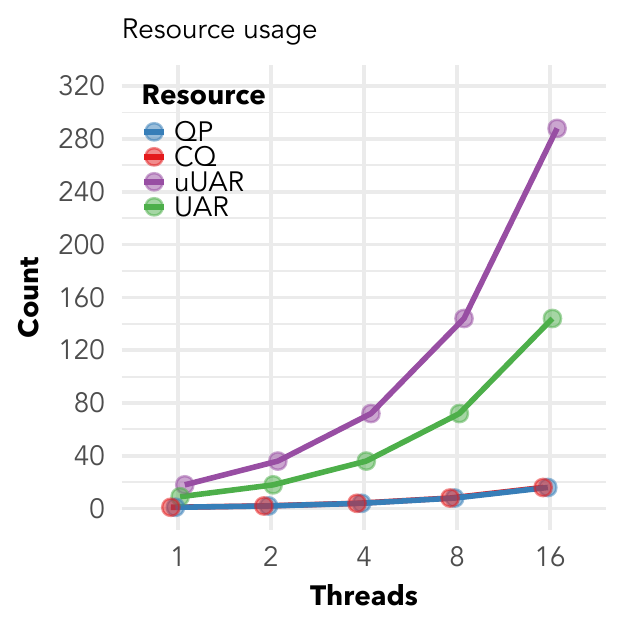}
	\end{minipage}
	\vspace{-2em}
	\caption{Scalability using a TD-assigned-QP per CTX per thread. Left: Throughput
		across features. Right: Resource usage.}
	\label{fig:scalability}
	\vspace{-1.5em}
\end{figure}

\figref{fig:scalability} shows the scalability\footnote{The
	NIC is attached only to the first socket; cross-socket
	NIC behavior is out of the scope of this work.} of communication
throughput across features and communication resource usage of endpoints created with
one TD-assigned QP per context per thread for 2-byte RDMA writes.
We observe that the number of QPs and CQs is an identity function of the
number of threads and increases their memory consumption
from 89 KB with one thread to 1.39 MB with 16 threads. The
usage of UARs and uUARs also grows by a factor of 9 and 18,
respectively. The reason is that each CTX containing one
TD allocates 9 UARs and each
UAR consist of two uUARs.

\section{Resource-Sharing Analysis}
\label{sec:analysis}

\begin{figure*}[htbp]
  \centering
  \begin{minipage}[t]{0.26\textwidth}
    \centering
    \includegraphics[width=\textwidth]{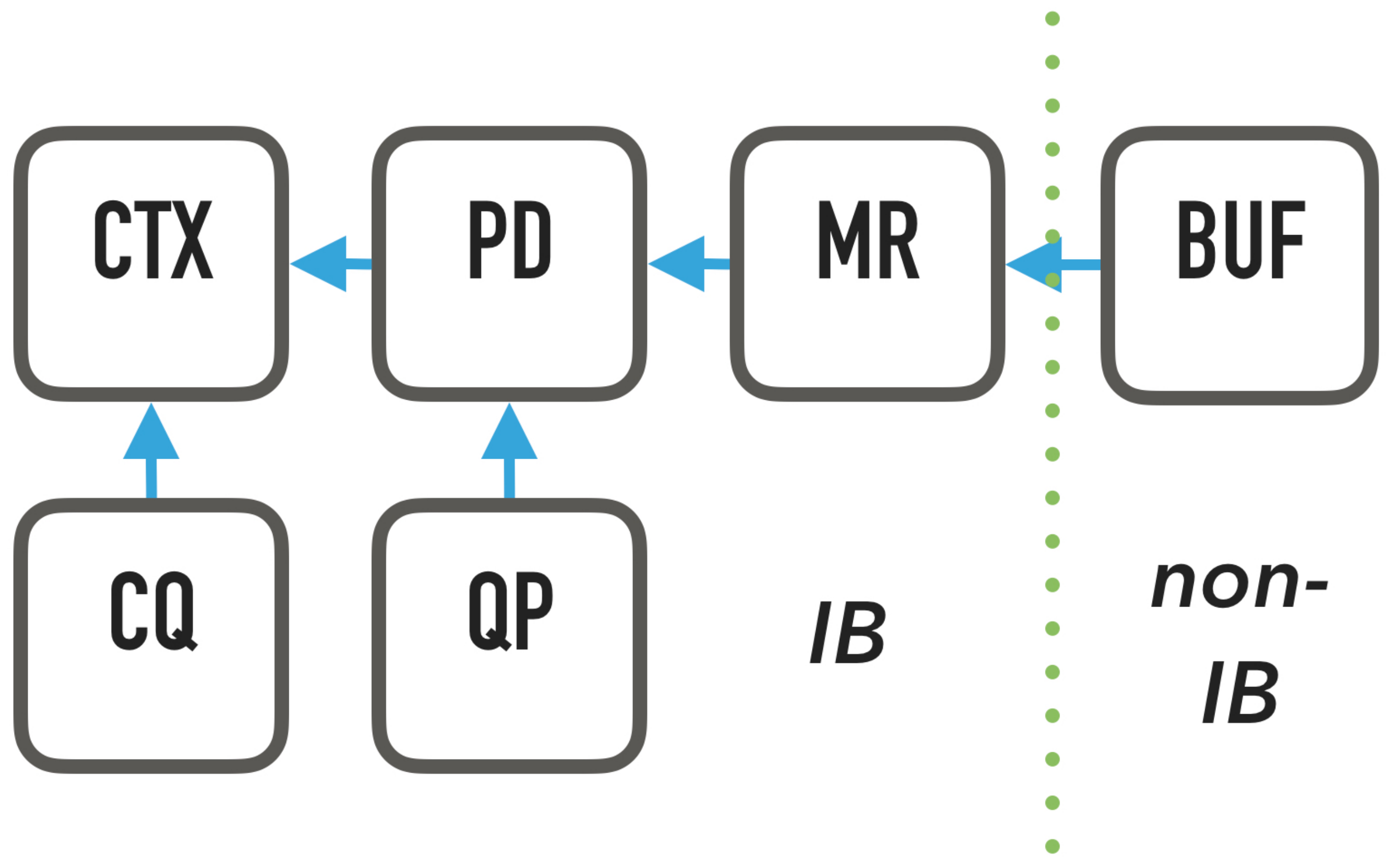}
    \par(a)
  \end{minipage}
  \begin{minipage}[t]{0.70\textwidth}
    \centering
    \includegraphics[width=\textwidth]{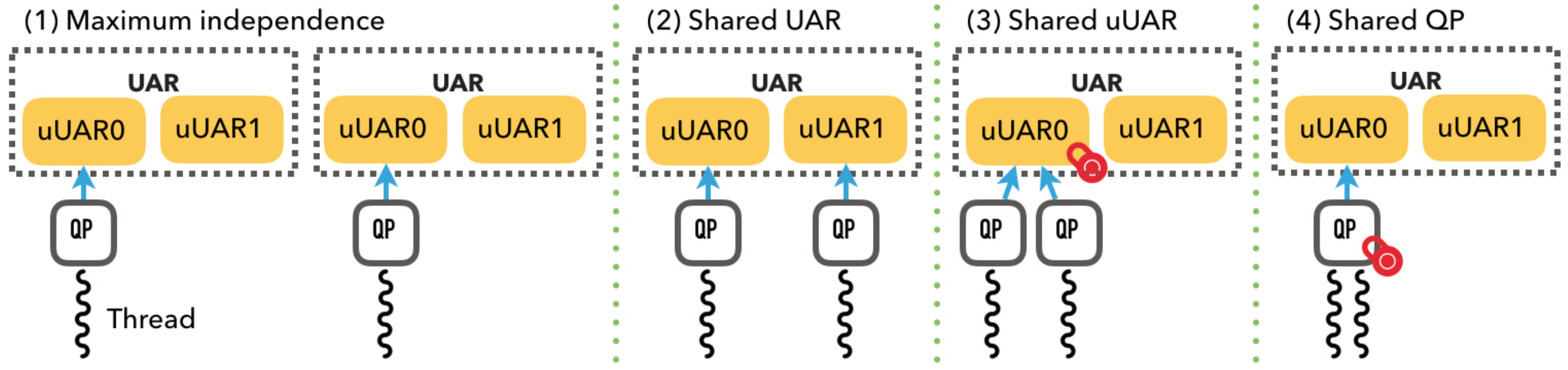}
    \par(b)
  \end{minipage}
  \caption{(a) Hierarchical relation between the various Verbs
    resources (the arrow points to the parent); each resource can have multiple children but only one parent. (b) Four levels of thread-to-uUAR
    mapping in mlx5 between independent threads.}
  \label{fig:sharinganalysis}
    \vspace{-1.75em}
\end{figure*}

From an analytical perspective, a thread can map to
the hardware resources in four possible ways.
 \figref{fig:sharinganalysis}(b)
demonstrates the four ways described below.
 
\begin{enumerate}
\item \emph{Maximum independence} -- There is no sharing of any
hardware resource between the threads; each is
assigned to its own UAR page (used in MPI everywhere). 

\item \emph{Shared UAR} -- The threads are assigned to
distinct uUARs sharing the same UAR page (mlx5 default
for multiple TDs described in \secref{sec:ctxsharing}).  

\item \emph{Shared uUAR} -- Although the threads have
their own QPs, the distinct QPs share the same uUAR
(medium-latency uUARs in ~\cite{scalable}). A lock is
needed on the shared uUAR for concurrent \blueflame\ writes. 

\item \emph{Shared QP} -- The threads share the same QP
(used in state-of-the-art MPI+threads), in which case a
lock on the QP is needed for concurrent device WQE
preparation. The lock on the QP also protects concurrent
\blueflame\ writes on the uUAR since the lock is released
only after a \blueflame\ write.
\end{enumerate}

Sharing software and hardware communication resources at
different levels improves resource efficiency but can hurt
throughput. Below, we explore the tradeoff space between resource
efficiency and communication throughput from the perspective of
the Mellanox IB user while considering the various IB
features described in \secref{sec:ibfeatures}. The user allocates
and interacts with the communication resources
through the IB
resources shown in \figref{fig:sharinganalysis}(a). Each of those
objects represents a level of sharing between threads.
Hence, we analyze the impact of sharing each IB resource on
performance and resource usage. We verify our analyses for 16
threads using the setup described in \secref{sec:setup}.

In the figures below, x-way sharing means the resource of interest
is being shared x ways. For example, 8-way sharing means
the resource is shared between 8 threads (two instances of
the shared resource). Moreover, we are interested in the change in
throughput with increasing sharing rather than the absolute
throughput obtained by using certain features.

Starting with \naive\ endpoints---each thread driving its own
set of resources using a TD-assigned QP---we move down each level of IB resource
sharing according to the hierarchical relation shown in
\figref{fig:sharinganalysis}(a). \figref{fig:scalability} shows the
performance and resource usage of this approach for 16 threads.

\subsection{Memory Buffer Sharing}
\label{sec:bufsharing}

\begin{figure}[htbp]
	\centering
	\begin{minipage}[t]{0.24\textwidth}
		\centering
		\includegraphics[width=\textwidth]{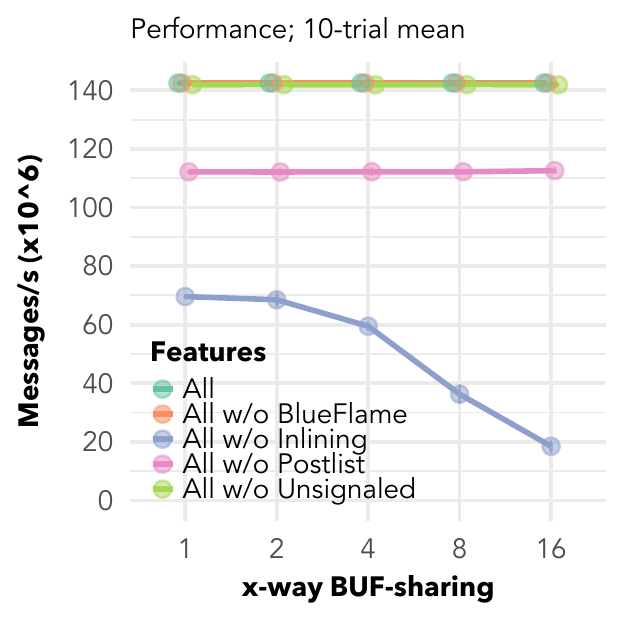}
	\end{minipage}
	\begin{minipage}[t]{0.24\textwidth}
		\centering
		\includegraphics[width=\textwidth]{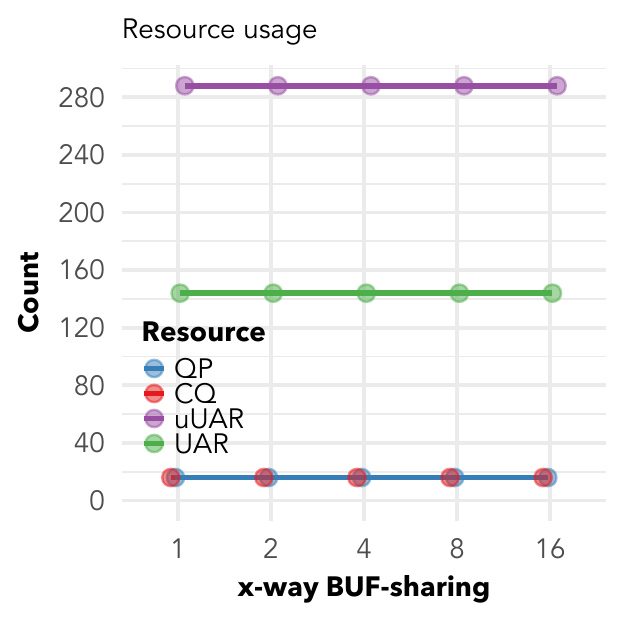}
	\end{minipage}
	\vspace{-2em}
	\caption{Message rate (left) and communication resource usage (right) with increasing
		BUF sharing across 16 threads.}
	\label{fig:bufsharing}
	      \vspace{-1.5em}
\end{figure}

The highest level of sharing is the non-IB resource: memory
buffer. We define the BUF to be the pointer to the
payload of the message. If the payload size is small enough, it can be inlined within
the WQE; that is, the CPU will read it. By default, the maximum message size that can
be inlined on ConnectX-4 exposed through Verbs is 60 bytes. Therefore,
for any larger message, the NIC must DMA-read the
payload.

\noindent \emph{\textbf{Performance.}} When the CPU reads
the payload, sharing this BUF between the threads is
safe since concurrent reads to the same memory
location in a CPU are harmless.
When the NIC reads the payload, however, its TLB design is important since a
virtual-to-physical address translation is imperative for the DMA
read. The NIC typically has a multirail TLB design that handles
multiple transactions in parallel in order to sustain the high speed
of the NIC's ASIC. The load is distributed across the TLBs by using a
hash function. If this hash function is based on the cache line,
concurrent DMA reads to the same cache line will hit the same
translation engine, serializing the reads. With a shared BUF, the WQEs
of multiple threads would point to the same cache line, serializing
the DMA reads.

\begin{figure}[htbp]
    \centering
    \begin{minipage}[t]{0.24\textwidth}
        \centering
        \includegraphics[width=\textwidth]{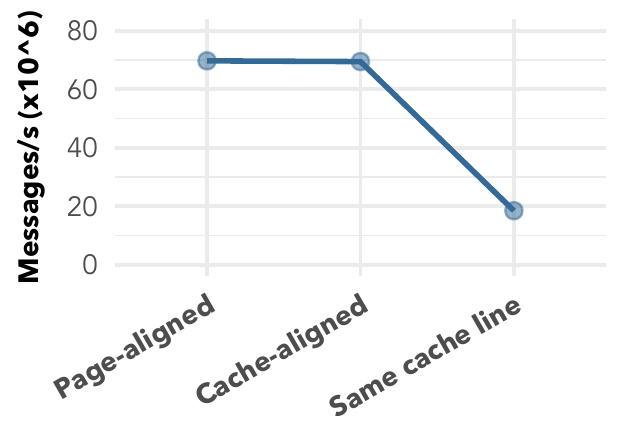}
        \par(a)
    \end{minipage}
    \begin{minipage}[t]{0.24\textwidth}
        \centering
        \includegraphics[width=\textwidth]{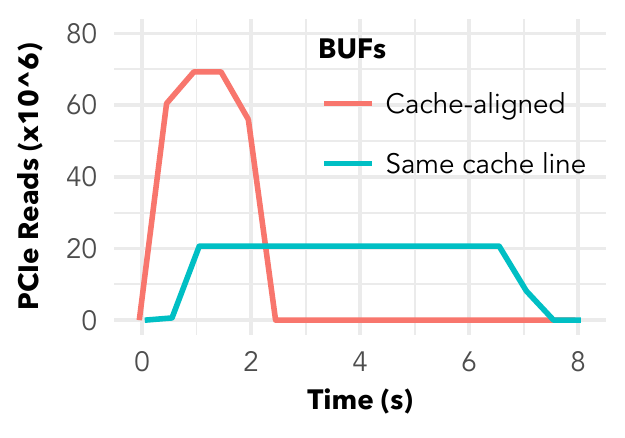}
        \par(b)
    \end{minipage}
    \caption{Effects on (a) message rate and (b) PCIe reads with and
        without cache-aligned buffers.}
    \label{fig:bufsharingexp}
      \vspace{-1.5em}
\end{figure}

\figref{fig:bufsharing} indeed shows that the throughput
decreases with increasing BUF sharing without \inlining\, that is, when the NIC reads the payload.
To further validate our analysis, \figref{fig:bufsharingexp}(a) shows that independent 2-byte buffers without
64-byte cache alignment also hurt performance since
all 16 buffers are on the same cache line.
While the total number of PCIe reads
(measured using PMU tools~\cite{pmu_tools}) with and without
cache alignment is equal,
\figref{fig:bufsharingexp}(b) shows that the rate of these PCIe
reads is much slower when the buffers are on the same cache
line. 

\noindent \emph{\textbf{Resource usage.}} The BUF is a non-IB
resource. Hence, it does not affect the usage of any of the
communication resources, as we can see in
\figref{fig:bufsharing}.

\subsection{Device Context Sharing}
\label{sec:ctxsharing}

We note that the Verbs user gets maximally independent (level 1 in \figref{fig:sharinganalysis}(b)) paths
without CTX sharing since the QPs naturally get assigned to uUARs
on different UARs. Within a shared CTX, however, the user has no way to explicitly request
maximally independent paths for multiple QPs. When the user creates multiple TDs, the
mlx5 provider can assign the threads to a uUAR using either the first
or the second level of sharing, as shown in
\figref{fig:sharinganalysis}(b). Currently, the mlx5 provider is hardcoded to use
the second level of sharing for multiple TDs, restricting the user
from creating maximally independent QPs within a CTX. More abstractly,
the Verbs users today have no way to request a sharing level for
the QPs/TDs they create. The number of levels of sharing is provider
specific.

To overcome this Verbs design limitation, we propose a variable,
\texttt{sharing}, in the TD initialization attributes (\texttt{struct
  ibv\_td\_init\_attr}) that are passed during TD creation. The higher
the value of \texttt{sharing}, the higher is the amount of hardware
resource sharing between multiple TDs. A \texttt{sharing} value of $1$
refers to maximally independent paths. In mlx5, only two levels of
sharing exist for TDs, corresponding to (1) and (2) in
\figref{fig:sharinganalysis}(b).

Note that the second uUAR of the UAR dedicated to a maximally
independent TD is wasted. Since the number of hardware resources is
limited, the user can request only a certain maximum number of
independent hardware resources within a CTX. This would be half of the
maximum number of UARs the user can dynamically allocate using TDs. In
mlx5, the maximum number of maximally independent paths is 256.

\begin{figure}[htbp]
	\centering
	\begin{minipage}[t]{0.24\textwidth}
		\centering
		\includegraphics[width=\textwidth]{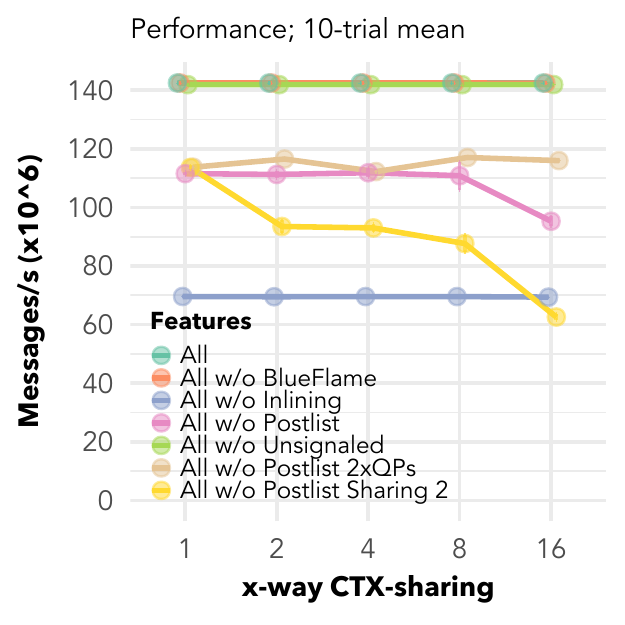}
	\end{minipage}
	\begin{minipage}[t]{0.24\textwidth}
		\centering
		\includegraphics[width=\textwidth]{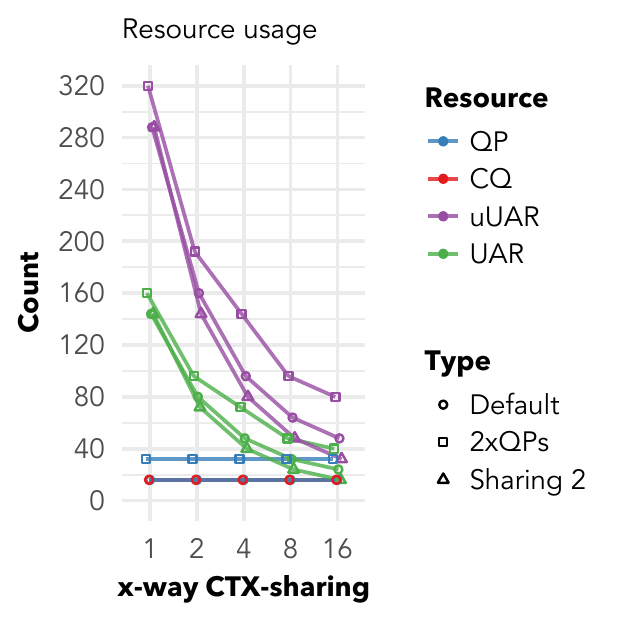}
	\end{minipage}
	\vspace{-2em}
	\caption{Message rate (left) and communication resource usage (right) with increasing
		CTX sharing across 16 threads.}
	\label{fig:ctxsharing}
	\vspace{-1.5em}
\end{figure}

Furthermore, we note that when the user assigns a QP to a TD, the
lock on the QP is still obtained. The mlx5 provider currently removes
only the lock on the uUAR that the TD
is assigned to. Since the user guarantees no concurrent access from
multiple threads to a QP assigned to a TD, the lock on the QP itself
can be disabled. We optimize the mlx5 provider for this case~\cite{mlx5lockcontrol}.

\noindent \emph{\textbf{Performance.}} For maximally independent
threads, sharing the CTX should not affect performance since we
emulate the thread-to-uUAR mapping in the MPI-everywhere
model. Sharing a CTX with the second level of sharing
between threads could hurt performance---the uUARs on the same
UAR could be sharing the same set of the NIC's registers, negatively
impacting throughput. Additionally, the CPU architecture's implementation of flushing write
combining memory can impact performance in the second level of
sharing since the memory attribute of the uUARs is set at the 
page-level granularity by using the Page Attribute Table (PAT)~\cite{pat}.

\figref{fig:ctxsharing} shows that sharing the CTX
does not hurt performance except when we
do not use \postlist\, that is, when we use \blueflame\ writes. 
For example, we notice a 1.15x drop in performance going
from 8-way to 16-way CTX sharing even with maximally independent TDs.
While the engineers at Mellanox are able to
reproduce this drop even on the newer ConnectX-5,
the cause for the drop is unknown. We discovered that creating
twice the number of maximally independent TDs but using only
half of them (even or odd ones) can eliminate this drop, as
seen in the ``All w/o Postlist 2xQPs" line. Additionally,
from the ``All w/o Postlist Sharing 2" line,
we can see the harmful effects of sharing a UAR when the
mlx5 provider is hardcoded to use the second sharing level for
assigning TDs within a shared CTX to uUARs.

While this evaluation validates the
need for maximally independent paths, it does not explain
the decline in throughput when there are concurrent \blueflame\
writes to distinct uUARs sharing the same UAR page. Finding
the precise reason for this behavior is hard since the
hardware-software interaction is dependent on the aforementioned proprietary
technologies.

\noindent \emph{\textbf{Resource usage.}} Sharing the CTX is critical
for hardware resource usage, as seen in \figref{fig:ctxsharing}. The reason is
that a maximally independent TD within a shared CTX adds only
1 UAR as opposed to 9 UARs when it is created within its own
independent CTX. Also, the 16 uUARs
and 8 UARs statically allocated by the mlx5 provider during CTX
creation (see \secref{sec:ibresources}) are wasted only once.
Nonetheless, maximally independent TDs will waste one uUAR
per thread. While sharing the CTX does not impact QP and CQ usage, it does reduce
the overall memory consumption. For example, when shared between 16 threads, it can reduce the
overall memory consumption by 9x (from 5.15 MB to 0.35 MB). 

\begin{figure}[htbp]
	\centering
	\begin{minipage}[t]{0.24\textwidth}
		\centering
		\includegraphics[width=\textwidth]{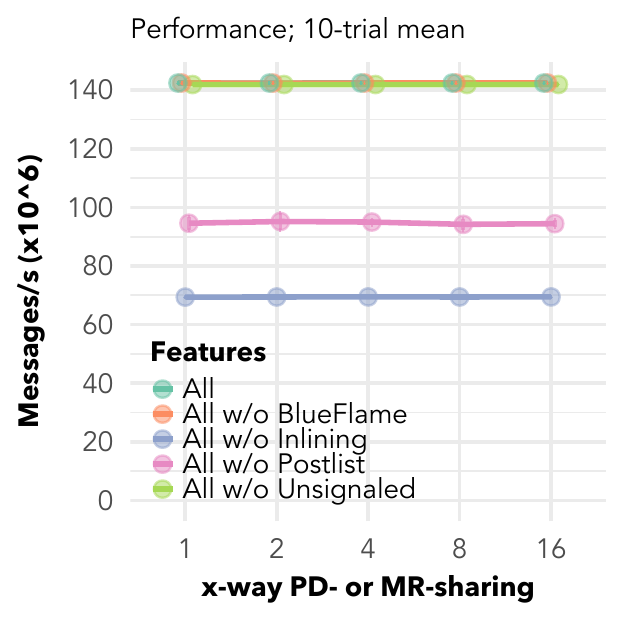}
	\end{minipage}
	\begin{minipage}[t]{0.24\textwidth}
		\centering
		\includegraphics[width=\textwidth]{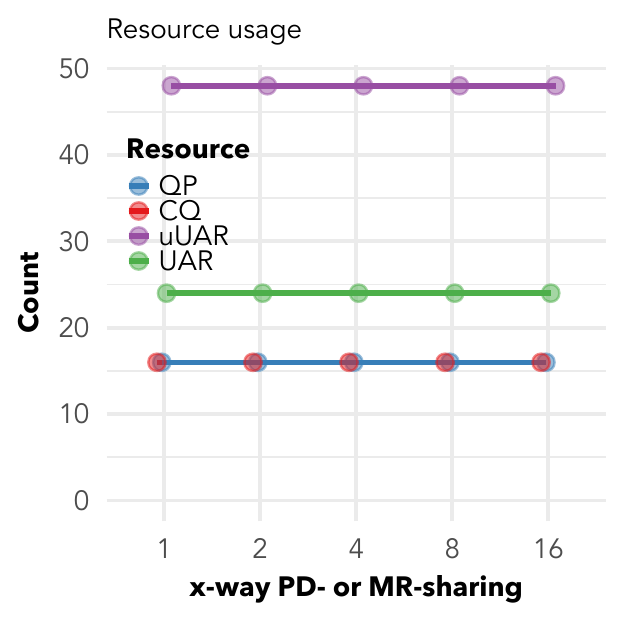}
	\end{minipage}
	\vspace{-2em}
	\caption{Message rate (left) and communication resource usage (right) with increasing
		PD or MR sharing across 16 threads.}
	\label{fig:pdnmrsharing}
	\vspace{-1.5em}
\end{figure}

Creating twice as many TDs (``2xQPs" in \figref{fig:ctxsharing})
increases resource usage since each the extra 16 maximally independent TDs allocates their
own QP and UAR. The second level of sharing
that mlx5 is hardcoded to use consumes 2x fewer 
UARs than do maximally independent TDs.

\subsection{Protection Domain Sharing}
\label{sec:pdsharing}

The protection domain is just a means of isolating a collection of IB
resources. Resources contained under different PDs
cannot interact with each other.

\noindent \emph{\textbf{Performance.}} The software PD object is not
accessed on the critical data-path; the protection checks occur in the
NIC. Hence, from a performance perspective, sharing a PD between
multiple threads would be harmless, as observed in
\figref{fig:pdnmrsharing}.

\noindent \emph{\textbf{Resource usage.}} The PD does not impact
the usage of any of the communication resources, as we can see
in \figref{fig:pdnmrsharing}. The uUAR and UAR values reflect
those of one CTX since
the PD can be shared only within a CTX.

\subsection{Memory Region Sharing}
\label{sec:mrsharing}

The MR is an object that pins memory in the
virtual address space of the user with the OS 
and prepares it for DMA accesses from the NIC.

\noindent \emph{\textbf{Performance.}} Sharing the MR between
threads will have no impact on performance since the MR is just an
object that points to a registered memory region. The MR may span
multiple contiguous BUFs. Sharing an MR
containing only one BUF means that the threads are sharing the BUF as well,
which implies the same effects of BUF sharing. \figref{fig:pdnmrsharing} confirms
that sharing the MR does not affect performance as long as the
threads have independent cache-aligned buffers.

\noindent \emph{\textbf{Resource usage.}} The MR does not
control the allocation of any of the communication resources. Hence,
sharing it will have no impact, as we can see in \figref{fig:pdnmrsharing}.

\ignore{
\begin{figure}[htbp]
    \centering
    \begin{minipage}[t]{0.24\textwidth}
        \centering
        \includegraphics[width=\textwidth]{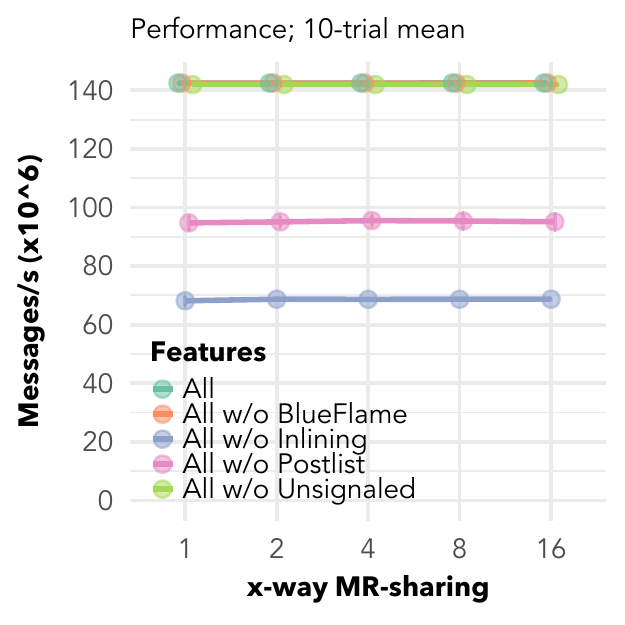}
    \end{minipage}
    \begin{minipage}[t]{0.24\textwidth}
        \centering
        \includegraphics[width=\textwidth]{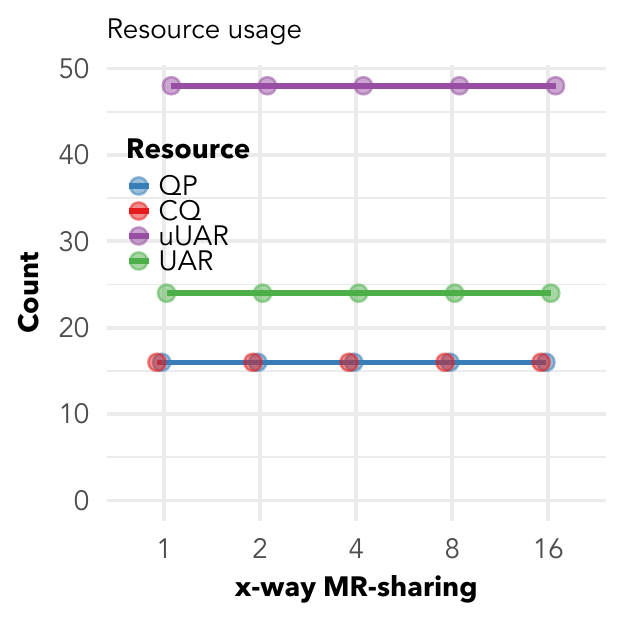}
    \end{minipage}
    \vspace{-2em}
    \caption{Message rate (left) and communication resource usage (right) with increasing
        MR sharing across 16 threads.}
    \label{fig:mrsharing}
\end{figure}
}

\begin{figure}[htbp]
	\centering
	\begin{minipage}[t]{0.24\textwidth}
		\centering
		\includegraphics[width=\textwidth]{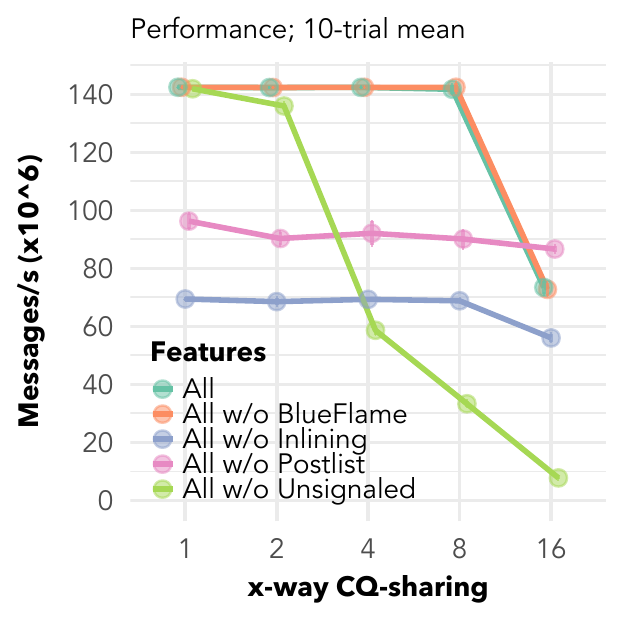}
	\end{minipage}
	\begin{minipage}[t]{0.24\textwidth}
		\centering
		\includegraphics[width=\textwidth]{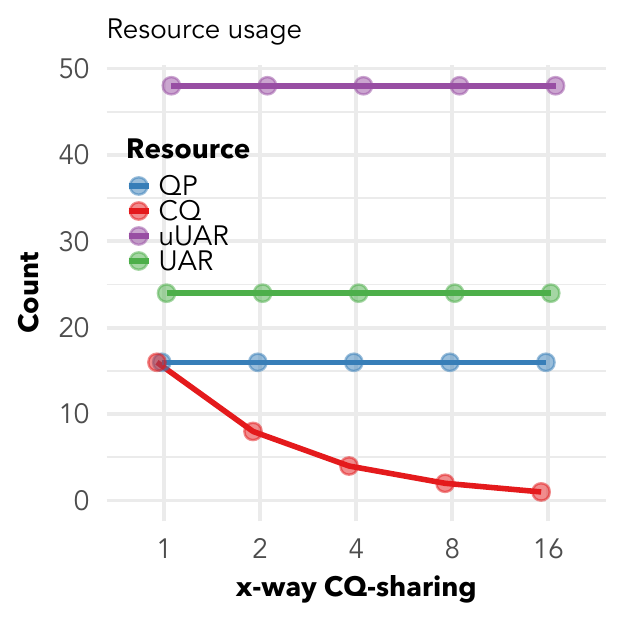}
	\end{minipage}
	\vspace{-2em}
	\caption{Message rate (left) and communication resource usage (right) with increasing
		CQ sharing across 16 threads.}
	\label{fig:cqsharing}
	\vspace{-1em}
\end{figure}

\subsection{Completion Queue Sharing}
\label{sec:cqsharing}

The Verbs user can map multiple QPs to the same CQ, allowing
for CQ-sharing between threads. In a latency-bound application, the user actively polls the CQ on the
critical data-path to confirm progress in communication. 

\begin{figure}[htbp]
	\begin{center}
		\includegraphics[width=0.49\textwidth]{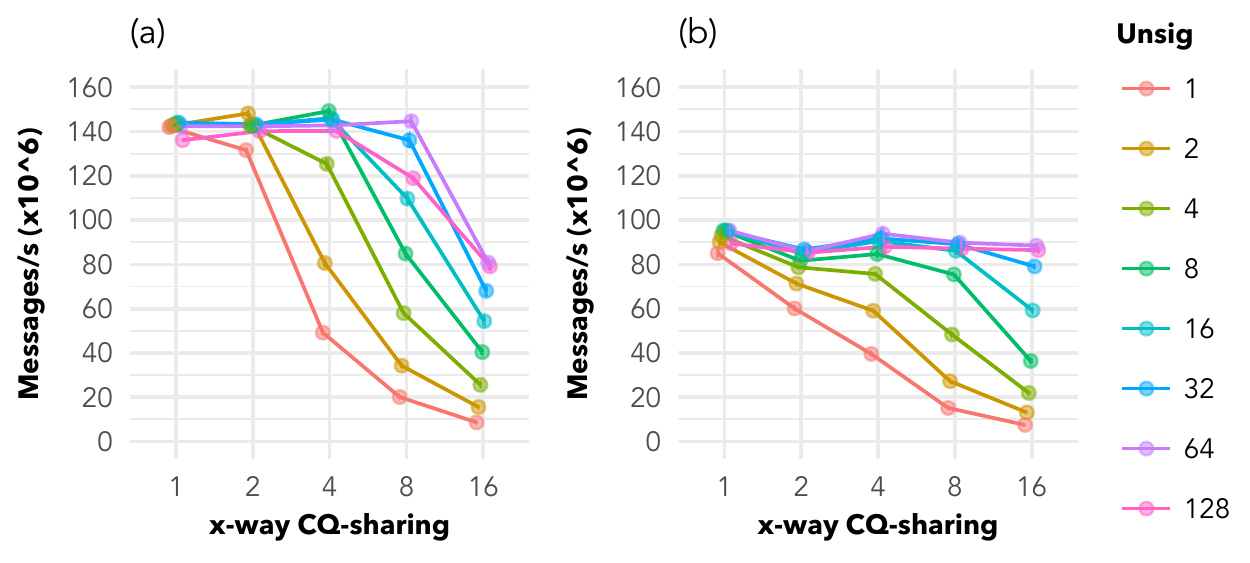}
	\end{center}
	\vspace{-1.5em}
	\caption{(a) Postlist size of 32, (b) Postlist size of 1.}
	\label{fig:varyingunsig}
	\vspace{-1.5em}
\end{figure}

\noindent \emph{\textbf{Performance.}} The CQ has a lock that a thread
will acquire before polling it. Hence, the threads sharing a CQ will
contend on its lock. Additionally, if QP $i$ and QP $j$ share a CQ,
then thread $i$ driving QP $i$ can read QP $j$'s completions. Hence,
the completion counter for any thread $i$ requires atomic updates.
Atomics and locks are obvious sources of contention when sharing
CQs between threads. \figref{fig:cqsharing} demonstrates
these hurtful effects of CQ sharing. The effects are most noticeable in
16-way sharing because there exists a tradeoff space between the
benefits of \unsigs\ and the overheads of CQ sharing.
\figref{fig:varyingunsig}(a) portrays this tradeoff space.
Lower values of \unsigs\ imply that the thread reads more completions
from the CQ than for higher values, translating to longer hold-time
of the shared CQ's lock. Thus, the impact of lock contention is
most visible in ``All w/o Unsignaled." For higher \unsig-values, we
see a drop only after a certain level of CQ sharing because the benefits of
\postlist\ outweigh the impact of contention. Removing
\postlist\ shows a linear decrease in throughput with increasing
contention in \figref{fig:varyingunsig}(b).

We note that even if the Verbs user can guarantee single-thread
access to a CQ, the standard CQ does not allow the user to disable
the lock on the CQ. The extended CQ, on the other hand, allows
the user to do so during CQ creation (\texttt{ibv\_create\_cq\_ex})
with the \texttt{IBV\_CREATE\_CQ\_ATTR\_SINGLE\_THREADED} flag.

\noindent \emph{\textbf{Resource usage.}} Sharing the CQ translates to
fewer circular buffers, and hence it reduces the memory consumption of the
completion communication resource. But it does not affect hardware
resource usage, as we can see in \figref{fig:cqsharing}. The uUAR and 
UAR usage shown corresponds to that of one CTX since a CQ can be shared
only within a CTX.

\subsection{Queue Pair Sharing}
\label{sec:qpsharing}

Ultimately, the user can choose to share the queue pair between
threads to achieve maximum resource efficiency. This is the case in
state-of-the-art MPI implementations.

\begin{figure}[htbp]
	\centering
	\begin{minipage}[t]{0.24\textwidth}
		\centering
		\includegraphics[width=\textwidth]{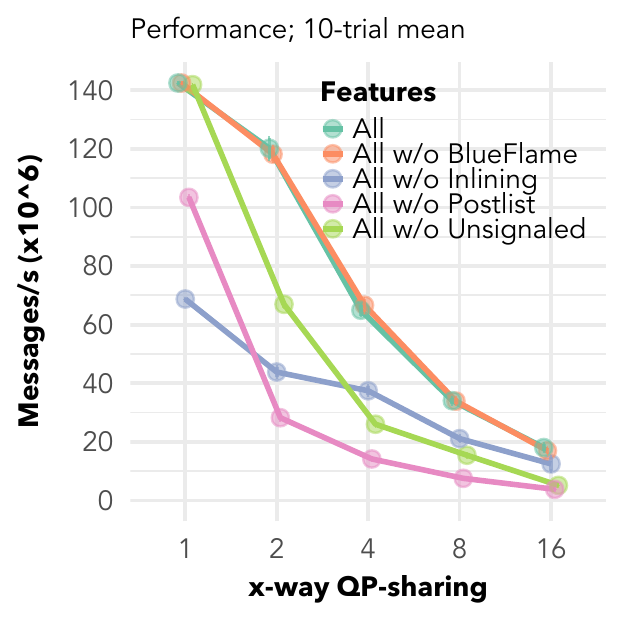}
	\end{minipage}
	\begin{minipage}[t]{0.24\textwidth}
		\centering
		\includegraphics[width=\textwidth]{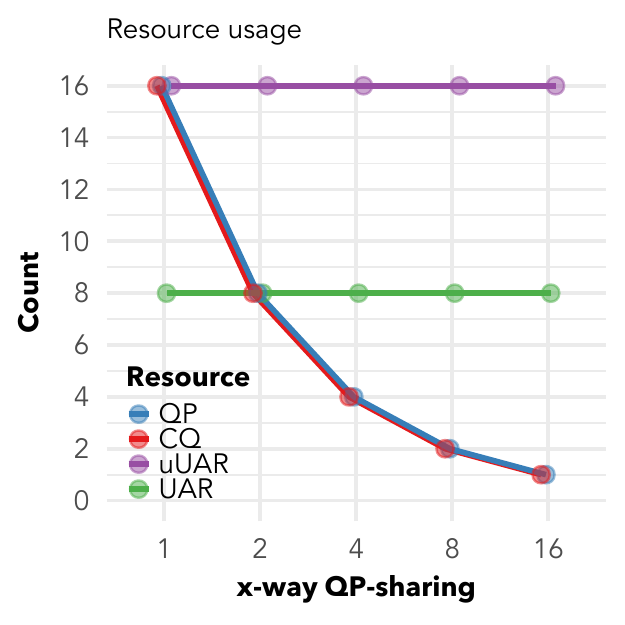}
	\end{minipage}
	\vspace{-2em}
	\caption{Message rate (left) and communication resource usage (right) with increasing
		QP sharing across 16 threads.}
	\label{fig:qpsharing}
	\vspace{-1.5em}
\end{figure}

\noindent \emph{\textbf{Performance.}} The QP has a lock that a thread
needs to acquire before posting on it. Hence, threads will contend on a
shared QP's lock. Additionally, the threads need to coordinate to post
on the finite QP-depth of the shared QP using atomic fetch and decrement 
the QP-depth value. These locks and atomics are sources of
contention when sharing QPs. Most important, the NIC's
parallel capabilities are not utilized with shared QPs since each QP
is assigned to only one hardware resource through which the messages
of multiple threads are serialized. \figref{fig:qpsharing}(a) shows the
expected decline in throughput with increasing QP-sharing. Removing
\postlist\ is more detrimental than removing \unsig\ because the
contention on the QP's lock without \postlist\ is higher.

\noindent \emph{\textbf{Resource usage.}} Sharing the QP means
fewer circular buffers for the WQEs and hence lower memory
consumption. It does not affect hardware resource usage, as we
can see in \figref{fig:qpsharing}(b). QP sharing reduces 
the number of both QPs and CQs, reducing the total memory
consumption of the software communication resources by 16x with
16-way sharing.

\noindent\emph{\textbf{Summary.}} Below are the lessons learned
from our analysis.
\begin{compactitem}
    \item Each thread must have its own cache-aligned buffer to prevent
    a performance drop.
    \item CTX-sharing is the most critical for the usage of
    hardware resources. With 16-way sharing, ``2xQPs" can achieve the same
    performance as independent CTXs using 80 uUARs instead of 288.
    If 20\% less performance is acceptable, we can use maximally
    independent TDs that use 6x fewer resources. If 50\% less performance
    is acceptable, we can use ``Sharing 2" that uses 9x fewer resources.
    \item Sharing the PD or the MR will not hurt performance, while keeping them 
    independent will not utilize any communication resource. 
    \item Only QP- and CQ-sharing affects the memory consumption of the software resources. 
    However, the reduction in memory usage by sharing them is not as critical as
    the consequent drop in performance. For example, 16-way sharing of the CQ
    improves memory usage by 1.1x but can result in an 18x drop in performance.
\end{compactitem}

\section{Designing Scalable Endpoints}
\label{sec:scalableepdesign}

Building on our
analysis, we define
the \emph{scalable endpoints} resource
sharing model that concretely categorizes
the design space of multiple communication
endpoints into six categories. Below we describe
the design of the initiation interface
in each category, state how the user can create it, discuss what
occurs internally in the IB stack, and
discuss its implications on performance and
resource usage.
For simplicity, we maintain
a separate CQ for each QP.

\noindent \textbf{\emph{MPI everywhere.}} This
category emulates the endpoint configuration
when multiple ranks run on a node. It
represents level 1 in
\figref{fig:sharinganalysis}(b). The user creates
this by creating a separate CTX for each thread,
each containing its own QP and CQ.
Within each CTX, the mlx5 driver assigns the QP
to a low-latency uUAR. Since each CTX contains
8 UARs, consecutive QPs naturally get assigned
to distinct UAR pages. The
performance of this category is the closest
to the best possible since there is no
sharing of resources. It is not the best since
the lock on the QP is still taken even though no other thread
contends for it. The resource usage of
this category is high: every CTX allocates 8 UARs.
Additionally, it is wasteful since only 1
of the 16 allocated uUARs is used
per thread. The
memory consumption increases linearly
with the number of threads since the
number of QPs and CQs is an identity
function of the number of threads.

\noindent \textbf{\emph{2xDynamic.}} This category also
represents
a 1-to-1 mapping between a
uUAR and a thread. Unlike MPI everywhere,
however, the user creates only one CTX for all the
threads and creates twice as many TD-assigned-QPs 
as threads. The threads use only the even or odd QPs.
The mlx5 provider dynamically
allocates a new UAR page for each TD and
assigns the first uUAR to
the TD, enabling a 1-to-1 mapping. 
This category delivers the best
performance. Since the number of QPs is twice
the number of threads, however, each thread wastes 1
dynamically allocated UAR, 3 uUARs, and 1
QP. The memory consumption
of QPs and CQs is twice that of MPI everywhere. The
statically allocated hardware resources are
wasted regardless of the number of threads.

\noindent \textbf{\emph{Dynamic.}} This category also
represents a 1-to-1 mapping between a uUAR
and a thread, but the number QPs equals the
number of threads. The user creates this
configuration similar to ``2xDynamic" by creating only as many QPs
as threads. 
According to
\secref{sec:ctxsharing}, this configuration
hurts communication throughput. In terms
of resource usage, however, only one uUAR is wasted
per thread. The 8 statically allocated UARs
are naturally wasted; none of the dynamically
allocated UARs are wasted. The memory consumption
of QPs and CQs is half of that in ``2xDynamic"
and same as MPI everywhere.

\noindent \textbf{\emph{Shared Dynamic.}} This category
represents level 2 in \figref{fig:sharinganalysis}(b). The user
creates this configuration using a shared CTX,
similar to the way in ``Dynamic," but assigns
each QP to a TD with the second level of sharing.
The mlx5 driver will dynamically allocate UARs
only for the even TDs and map the even TDs
to the first uUAR and the odd TDs to the second
uUAR of the allocated UAR.
According to
\secref{sec:ctxsharing}, sharing the UAR will hurt
performance. The hardware resource usage is less
than with ``Dynamic" since only half as many UARs and
uUARs as threads are allocated. Apart from the 8
statically allocated UARs and uUARs, none of the
dynamically allocated resources are wasted. The
memory consumption of QPs and CQs is
equivalent to that of ``Dynamic."

\noindent\textbf{\emph{Static.}} The user uses the statically allocated resources
within a CTX, resulting in a many-to-one mapping
between the threads and uUARs (and UARs). To do so, the user 
simply creates a QP for each thread within a
shared CTX without any TDs. The final
state of the mapping for a given number of QPs
is dependent on the driver's assignment policy. In mlx5, with
16 QPs, we end up with a
combination of the second and third level of sharing
in \figref{fig:sharinganalysis}(b)---the $5$th and 
$16$th QP are mapped to the same uUAR (third
level), while the others are mapped to the rest of
the uUARs using the second level of sharing.
The hardware resource usage is the number of statically
allocated resources. Resources are
wasted only when the number of threads
is less than 16. The memory consumption
is equivalent to that of ``Dynamic."

\noindent \textbf{\emph{MPI+threads.}} This
category represents level
4 in \figref{fig:sharinganalysis}(b). The
user creates this by creating only 1 CTX,
1 QP, and 1 CQ. The mlx5 driver assigns
the one QP to a low-latency uUAR. The
performance of this category is the worst
possible since the communication of all the
threads is bottlenecked through one QP.
The resource usage of this category is not
a function of the number of threads and
hence is the best possible. All threads
allocate only 8 UARs, 16 UARs, 1
QP, and 1 CQ.

Note that the CQ can be shared in any manner in the above
categories and its impact is orthogonal to
the effects of the initiation interface. 

\section{Evaluating Scalable Endpoints}
\label{sec:scalableepeval}

We evaluate the performance
and resource usage of scalable endpoints
described in \secref{sec:scalableepdesign} on two benchmarks, namely, global array and 5-point stencil
on our two-node evaluation setup\footnote{
	Thread domains are supported only kernel 4.16
	onward; the latest stable kernel was 4.17.2, hence,
	only two nodes since the combination of a
	mlx5 device along with the latest stable kernel was a rarity.}.
We limit our evaluation
to conservative application semantics---those
that do not allow \postlist\ and \unsigs\ and
focus on \blueflame\ writes instead of \doorbells\
since they are latency oriented.

\noindent\emph{\textbf{Global array benchmark.}}
The pattern of fetching and writing tiles from
and to a global array is at the core of many scientific
applications such as NWChem~\cite{valiev2010nwchem},
which constitutes a multidimensional
double-precision matrix multiply (DGEMM).


We implement a DGEMM benchmark ($A \times B = C$),
where the global matrices $A$, $B$, and $C$ reside
on a server node and a client node performs
the DGEMM using Verbs for internode communication.
We design the benchmark such that all the QPs share
the same PD but each has three BUFs and three MRs---one
for each of the three tiles from A, B, and C.

\begin{figure}[htbp]
	\centering
	\begin{minipage}[t]{0.24\textwidth}
		\centering
		\includegraphics[width=\textwidth]{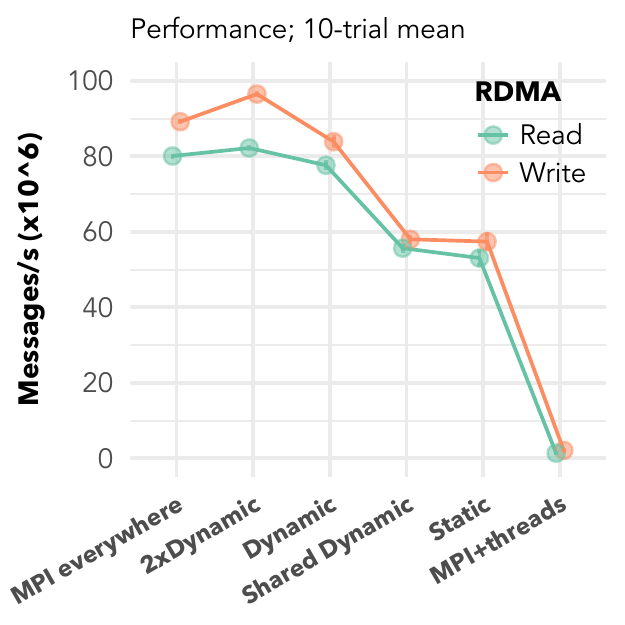}
	\end{minipage}
	\begin{minipage}[t]{0.24\textwidth}
		\centering
		\includegraphics[width=\textwidth]{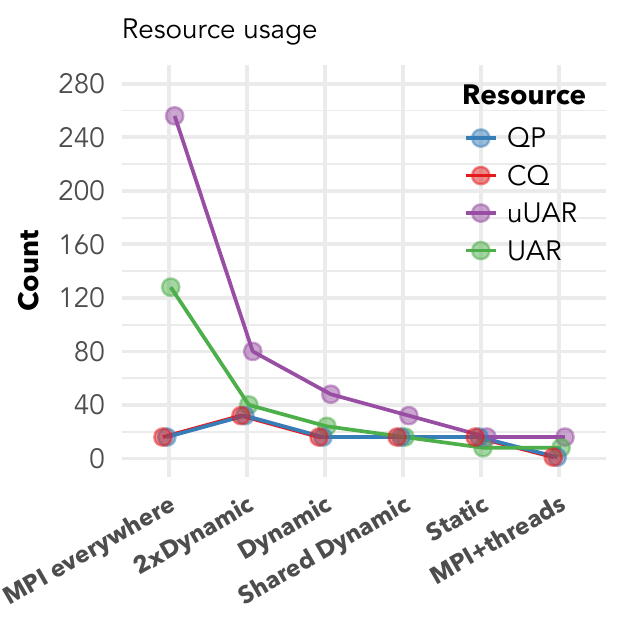}
	\end{minipage}
	\vspace{-2em}
	\caption{Scalable endpoints for the global array kernel with 16 threads.
		Left: Communication throughput. Right: Communication resource
		usage}
	\label{fig:gaeval}
	\vspace{-1.5em}
\end{figure}



\figref{fig:gaeval}
shows the performance and resource
usage of scalable communication endpoints
for 16 threads.
Performance
decreases with lower resource usage. For RDMA
writes, for example, we observe that using maximally
independent TDs with twice the number of QPs (2xDynamic)
gives us 108\% of the performance
of dedicated endpoints (MPI everywhere) while
using only 31.25\% as many hardware resources.
Maximally independent paths with as many QPs as threads (Dynamic) gives us
94\% of the performance of
MPI everywhere while using 18.75\% as many hardware
resources.
Sharing the UAR (Shared Dynamic)
gives us 65\% of the performance using 12.5\% of
the hardware resources.
Sharing the uUAR (Static) gives us
64\% of the performance using 6.25\% as many hardware
resources. We observe only a minimal drop in performance
in Static since only two threads share the
uUAR in Static; the rest share the UAR (see
\secref{sec:scalableepdesign}), and hence
we observe performance similar to Shared Dynamic.
Finally, sharing the QP results in only 3\% of the
performance while still using 6.25\%
as many hardware resources.

The memory consumption of QPs and CQs is the
same for all categories except 2xDynamic and
MPI+threads. While the number of QPs and CQs
in 2xDynamic is twice that of MPI everywhere,
the overall memory usage in the former is 3.27x
lower (1.64 MB vs 5.39 MB; see
\secref{sec:resources}) since MPI everywhere
has 16 CTXs while 2xDynamic has only one.
The memory consumption is the lowest in
MPI+threads with only one QP and one CQ.

\begin{figure}[htbp]
	\begin{center}
		\includegraphics[width=0.49\textwidth]{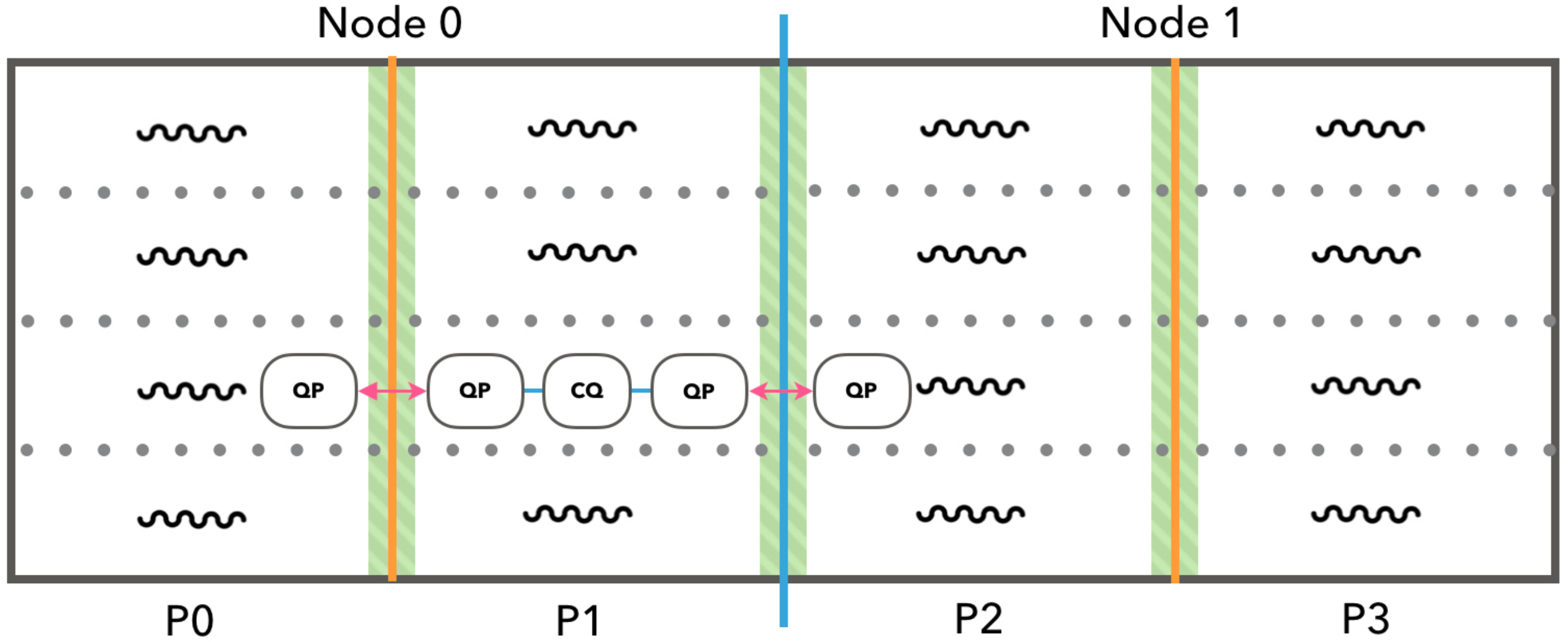}
	\end{center}
	\vspace{-1em}
	\caption{1-D partitioning of a grid for the 5-pt stencil between two nodes, four ranks
		(P0..P3), and four threads per rank showing the QP-CQ connection
		for each thread using one sample. The shaded regions are the halo regions.}
	\label{fig:stencildesign}
	\vspace{-1.5em}
\end{figure}
\begin{figure*}[htbp]
    \begin{center}
        \includegraphics[width=0.99\textwidth]{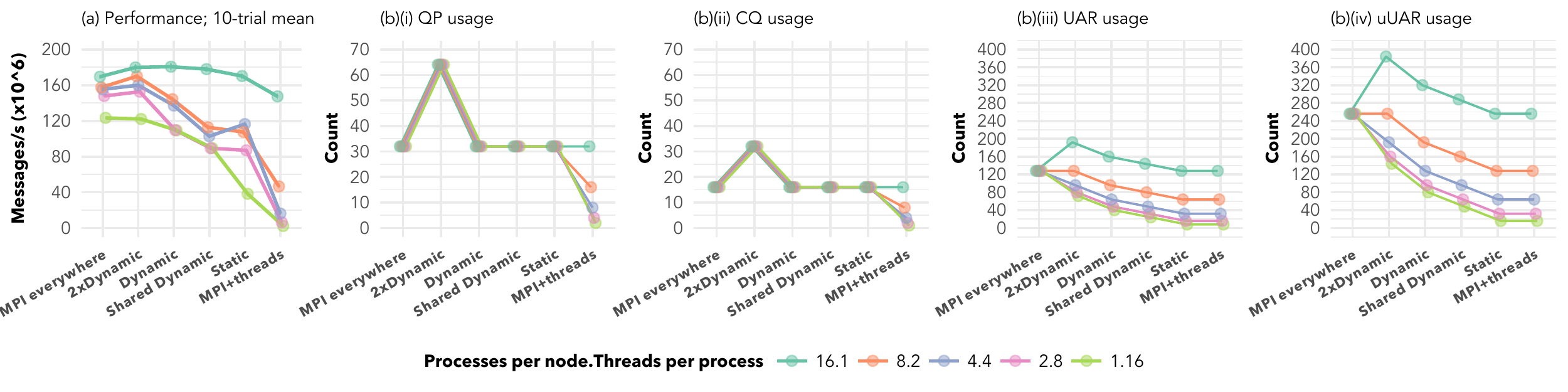}
    \end{center}
  	\vspace{-2em}
    \caption{(a) Message rate and (b) resource
        usage of (i) QP, (ii) CQ, (iii) UAR, and (iv) uUAR
        of 16 threads with scalable endpoints for a 5-pt stencil kernel.}
    \label{fig:stencileval}
\end{figure*}

\noindent\emph{\textbf{Stencil benchmark.}}
Stencil codes are at the heart of various application domains 
such as computational fluid dynamics, image processing, and partial
differential equation solvers. We evaluate scalable endpoints on a 5-pt stencil
benchmark with a 1D partitioning of the grid.
\figref{fig:stencildesign} shows the design of
our benchmark. We vary the number
of ranks per node and threads per rank such that
the total number of hardware threads engaged
is 16, the number of cores in a socket. Each
rank gets its tile from the grid, and each
thread gets a corresponding subtile. Each
thread requires two QPs, one for each of its
neighbors. We map the two QPs to
one CQ. Hence the number of QPs is twice the number
of CQs for all cases. \figref{fig:stencileval} shows
the performance,\footnote{The message rates are
    above 150 million, the maximum reported for
    ConnectX-4~\cite{x4record} since a majority of
    the halo exchanges are intranode. Intranode
    communication in InfiniBand still involves the NIC.} and resource usage of scalable
endpoints for the different hybrid scenarios.

For each category, a higher number of processes
performs better than a lower one. For MPI-everywhere
endpoints, for example, the fully hybrid approach
(1.16) performs 1.4x worse than the
processes-only approach (16.1). The
reason for this behavior is that the number of messages
with processes only is 16x higher, while 16 threads
per rank can exchange the halo only 7.67x faster than
with one thread per rank.

In the processes-only case, there is no resource
sharing since each
process has only one thread. 2xDynamic, Dynamic,
and Shared Dynamic achieve 106\% of MPI-everywhere's
performance because of the absence of the lock
on the QP. Static produces 100\% performance since the
lock on its QP exists. MPI+threads achieves
87\% of the performance even though there is no
contention between threads, because of the overhead
of atomics and additional branches associated with QP-sharing. In 16.1, the
number of QPs and CQs is the same for all cases
except for 2xDynamic, where they are 2x
higher. The hardware resource usage is higher in
2xDynamic, Dynamic, and Shared Dynamic since
they waste the statically allocated resources in each process,
unlike other categories.

For the hybrid cases, we observe a performance trend similar
to the global array kernel. 2xDynamic achieves 103\%
of the performance of MPI everywhere; and with increasing
resource-sharing, we improve resource usage but lose
performance. In the case of 4.4, of the eight QPs per CTX,
the fifth QP uses the first level of sharing in Static, resulting in eight
such QPs in total; hence, it performs better than Shared Dynamic
wherein all the QPs use only the second level of sharing. Similarly,
in 1.16, of the 32 QPs per CTX, 28 use the third level of sharing
in Static, hence performing worse than Shared Dynamic. For a given category, the
hardware resource usage is lower when the number of processes
is smaller since fewer processes mean fewer CTXs, and hence,
the total number of statically allocated resources is smaller.
Similarly, the number of QPs and CQs is the same for all hybrid
cases in all categories except in MPI+threads, where it is a function of the number of processes.

\section{Related Work}
\label{sec:related}
To the best of our knowledge, the resource-sharing analysis in this
paper to design a resource-sharing model at the low level of
interconnects is the first of its kind. The idea of multiple endpoints for
multinode programming models such as MPI and Unified Parallel C (UPC),
however, is not new. The research in this domain is motivated by the same
problem: loss in communication throughput in hybrid environments.

\noindent \textbf{MPI endpoints}. Dinan et al.~\cite{dinan2014enabling}
enable multiple communication endpoints by creating additional MPI ranks
that serve as the ``MPI endpoints." The threads within the MPI ranks then
map to the MPI endpoints, achieving the same configuration as MPI-everywhere
endpoints since each MPI endpoint has its CTX. However, they do not
consider the resource usage of their approach. Consequently, the 93.75\%
wastage of resources still holds with MPI endpoints. Our work explores the
tradeoff space between performance and resource usage instead of
providing one solution, allowing users to choose the best endpoint for their
needs.

\noindent \textbf{PAMI endpoints}. Tanase et al.~\cite{tanase2012network} implement
multiple endpoints for the IBM xlUPC runtime by assigning contexts
to UPC threads with a one-to-one mapping. While their work is a complete
solution, it does not demonstrate the indirect impact on the resource
usage. We show a holistic picture of the different mappings
between threads and hardware resources and discuss the
tradeoff between performance and resource usage for each
mapping.

\noindent \textbf{UPC endpoints}. Luo et al. ~\cite{luo2011multi}
implement network endpoints for the UPC runtime. However, their work
does not consider the mapping between the runtime's network endpoints
and the interconnects network resources. Consequently, their work
does not evaluate the hardware-resource utilization of their
implementation, which is an essential factor for understanding the
scalability of multiple communication endpoints. 

\section{Conclusions}
\label{sec:conclusion}

For a given number of hardware threads, state-of-the-art MPI
implementations either achieve maximum communication
throughput and waste 93.75\% of hardware resources using multiple
processes or achieve maximum resource efficiency and perform up to 7x
worse with multiple threads. In this work, we study the tradeoff space
between performance and resource usage that lies in between the
two extremes. We do so by first analyzing and evaluating in depth the
consequences of sharing network resources between independent
threads. In the process, we extend the existing Verbs design to allow
for maximally independent paths, for which case we also optimize the
mlx5 stack. As a result of our analysis, we describe \emph{scalable
communication endpoints}, an efficient resource sharing model for
multithreading scenarios at the lowest software level of interconnects.
Each category of the model reflects a performance level and its 
corresponding resource usage that users, such as MPICH, can use to
guide their creation of endpoints. The model's 2xDynamic endpoints, for
example, can achieve 108\% of the performance of the endpoints
in MPI-everywhere while using only 31.25\% as many resources.

\section*{Acknowledgment}

We thank Pavel Shamis from Arm Research, the members of
the linux-rdma@vger.kernel.org mailing list for their prompt responses,
Benjamin Allen and Kumar Kalyan from JLSE for their support, and the
members of the PMRS group
at ANL for their continuous feedback. This material is based upon work
supported by the U.S. Department of Energy, Office of Science, under
contract DE-AC02-06CH11357.

\bibliographystyle{abbrv}
\small
\bibliography{bib/refsused}

\appendices
\section{User Access Region}
\label{app:uar}

The User Access Region (UAR) is part of a mlx5 NIC's address space and
consists of UAR pages. Different pages allow the multiple processes and
threads to get isolated, protected, and independent direct access to the 
NIC. The UAR pages are mapped into the application's userspace during
CTX creation, allowing the user to bypass the kernel and directly write to the NIC. 

A mlx5 UAR page is 4 KB, and each UAR consists of four uUARs
(micro UARs). Only the first two are used for user operations; we refer
to them as data-path uUARs. The last two are used by the hardware for
executing priority control tasks~\cite{fastpathposting}. Each uUAR consists
of two equally sized buffers that are written to alternatively~\cite{mlxRPM}.
The first eight bytes of a buffer constitute the \doorbell\ register~\cite{mlxRPM}.
Atomically writing eight bytes to this register rings the \doorbell.

\begin{figure}[htbp]
    \begin{center}
        \includegraphics[width=0.49\textwidth]{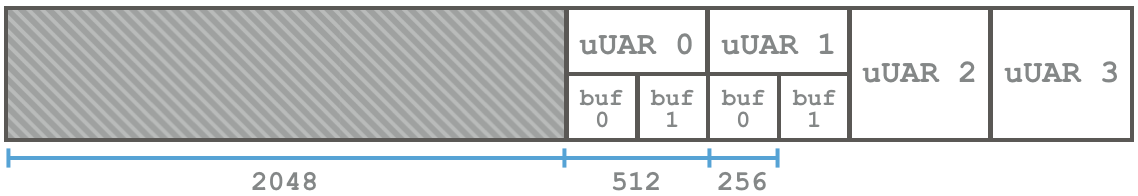}
    \end{center}
    \caption{4 KB mlx5 UAR page. The last two uUARs are used by the NIC.}
    \label{fig:mlx5uar}
\end{figure}

\section{mlx5's uUAR-to-QP Assignment Policy}
\label{app:uuartoqp}

When the Verbs user creates a CTX, the mlx5 driver statically allocates
a discrete number of UARs. By default, it allocates 8 UARs and 16
data-path uUARs. When the user creates QPs, the \texttt{mlx5\_ib} kernel
module assigns a uUAR to each QP. To guide this assignment, mlx5 categorizes
the statically allocated uUARs into different categories: the zeroth uUAR as
\emph{high latency}, a subset as \emph{low latency}, and the remaining as
\emph{medium latency}. By default, mlx5 categorizes four uUARs (uUAR12-15)
as low latency. Users can change this default using environment variables
that allow them to control the total number of statically allocated uUARs
(MLX5\_TOTAL\_UUARS) and categorize a subset of them (up to a maximum
of all but one) to be low-latency uUARs (MLX5\_NUM\_LOW\_LAT\_UUARS). 

Low-latency uUARs are called so because only one QP is assigned to such a
uUAR; thus the lock on the uUAR is disabled. The medium-latency uUARs
may be assigned to multiple QPs, and locks are needed to write to them. The
high-latency uUAR can also be assigned to multiple QPs but it allows only
atomic \doorbells\ and no \blueflame\ writes. Hence, it is not protected by a lock.

\begin{figure}[htbp]
    \begin{center}
        \includegraphics[width=0.49\textwidth]{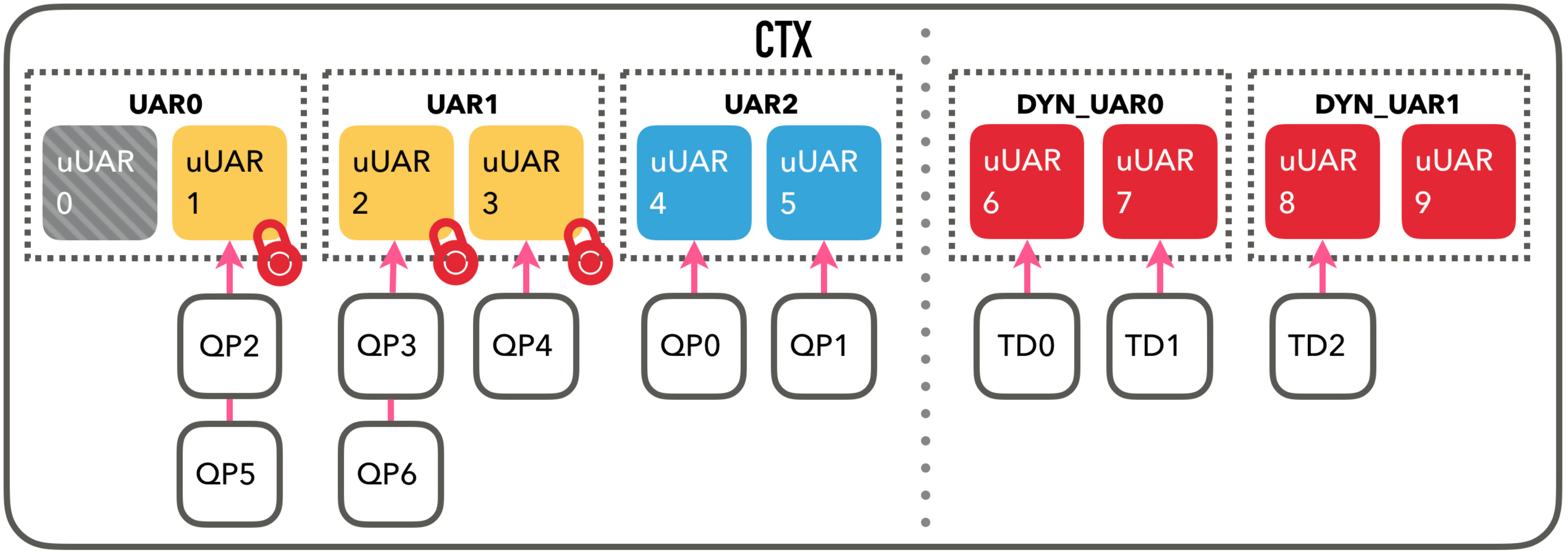}
    \end{center}
    \caption{Assigning seven QPs and three TDs to uUARs of a CTX
        containing six static uUARs, of which two are low-latency uUARs
        (blue). The high-latency uUAR is in grey, the medium-latency
        ones are in yellow, and the dynamically allocated ones are in red.}
    \label{fig:uuaralloc}
\end{figure}

\figref{fig:uuaralloc} portrays mlx5's uUAR-to-QP assignment
policy for an example CTX containing six static uUARs of which
two are low latency (uUAR4-5). Within a CTX, the QPs are first assigned
to the low-latency uUARs (QP0 and QP1). Once all the low-latency uUARs
are exhausted, the driver maps the next QPs to the
medium-latency uUARs in a round-robin fashion (QP2--QP6). The high-latency
uUAR is assigned to QPs only when the user declares the maximum allowed
number of uUARs to be low latency, in which case (not shown) all the QPs after
those assigned to the low-latency uUARs will map to the zeroth uUAR.

The mlx5 driver will \emph{dynamically allocate} a new UAR page if the user
creates a thread domain (TD). Every even TD will allocate a new UAR page;
every even-odd pair of TDs will map to the separate uUARs on the same UAR
page, as we can see for the three TDs in \figref{fig:uuaralloc}. All the QPs in a TD
will map to the uUAR associated with the TD; and since the user guarantees that
all the QPs assigned to a TD will be accessed only from one thread, mlx5 disables the
lock on the TD's uUAR. The maximum number of dynamically allocated UARs allowed
per CTX in mlx5 is 512.

\section{InfiniBand Mechanism}
\label{app:ibmech}

\begin{figure}[htbp]
    \begin{center}
        \includegraphics[width=0.49\textwidth]{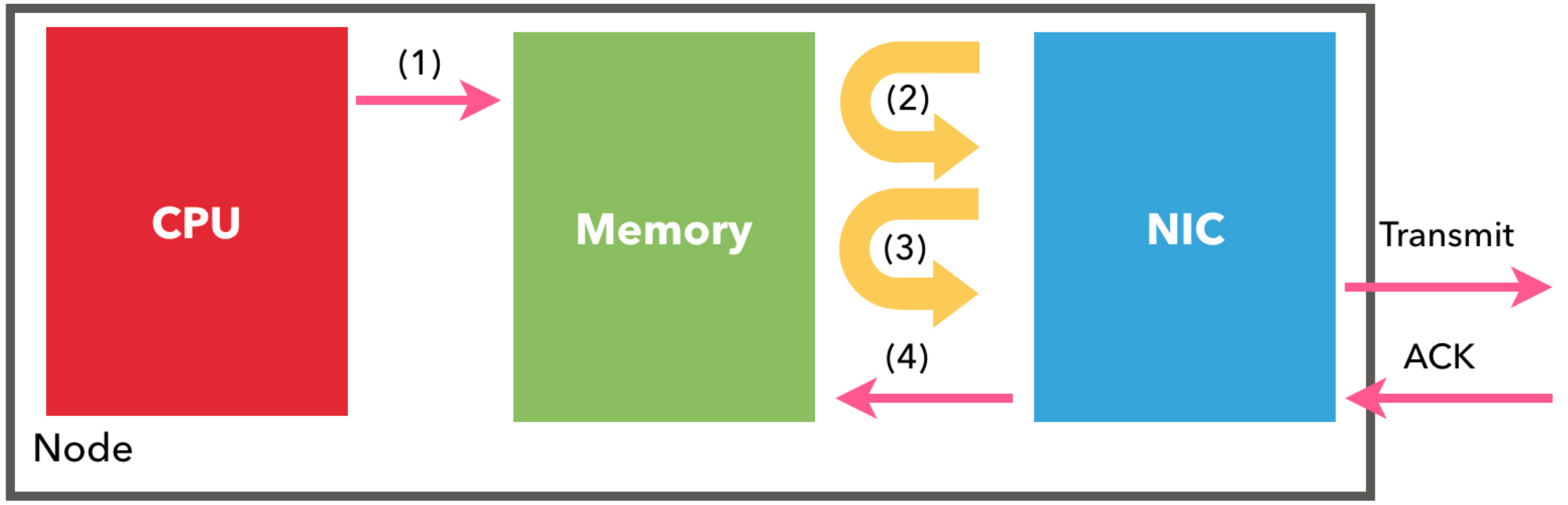}
    \end{center}
    \caption{IB mechanisms on sender node to send data over wire. Refer
        to ~\ref{sec:ibfeatures}, which describes each step in detail.}
    \label{fig:ibmech}
\end{figure}

To send the message over the InfiniBand network, the user posts a work queue
element (WQE) to a queue pair (QP) using an \texttt{ibv\_post\_send}. \figref{fig:ibmech}
portrays the series of coordinated operations between the CPU and the NIC that follow
to transmit a message and signal its completion. We describe them below.

\begin{enumerate}
    
\item Using an 8-byte atomic write (memory-mapped I/O) on the buffer of the uUAR
associated with the QP, the CPU first notifies the NIC that WQE is ready to be read.
This is called \emph{ringing the} \doorbell.

\item After the \doorbell\ ring, the NIC will fetch the WQE
using a DMA read. The WQE contains the virtual address
of the payload (stored in the WQE’s scatter/gather list).

\item The NIC will then fetch the WQE’s payload from a
registered memory region using another DMA read. Note
that the virtual address has to be translated to its physical
address before the NIC can DMA-read the data. The NIC
will then transmit the read data over the network.

\item Once the host NIC receives a hardware acknowledgment
from the receiver NIC, it will generate a CQE and DMA-write
it to the buffer (residing in host’s memory) of the
CQ associated with the QP. Latency-oriented users will
then poll this CQ to dequeue the CQE to ”make progress.”

\end{enumerate}

In summary, the critical data path of each \texttt{ibv\_post\_send}
entails one MMIO write, two DMA reads, and one DMA write.

\end{document}